\documentclass[a4paper,reqno,11pt]{amsart}

\usepackage[utf8]{inputenc}
\usepackage[T1]{fontenc}
\usepackage{lmodern}
\usepackage[UKenglish]{babel}
\usepackage[UKenglish]{isodate}
\usepackage{amsmath, amsfonts, amssymb, amsthm, mathtools, cancel} 
\usepackage{stmaryrd}
\usepackage{geometry} 
\usepackage{hyperref} 
\hypersetup{colorlinks = true}
\hypersetup{linktocpage} 
\usepackage[capitalise]{cleveref} 
\usepackage{dsfont}

\newcommand{\interval}[4]{\mathopen{#1}#2
	\mathclose{}\mathpunct{},#3
	\mathclose{#4}}

\newcommand{\intoc}[2]{\interval{(}{#1}{#2}{]}}
\newcommand{\intco}[2]{\interval{[}{#1}{#2}{)}}
\newcommand{\intoo}[2]{\interval{(}{#1}{#2}{)}}
\newcommand{\intinteger}[2]{\interval\llbracket{#1}{#2}\rrbracket}

\newcommand{\bigO}[1]{O\mathopen{}\left(#1\right)}

\newcommand{\abs}[1]{\left\lvert#1\right\rvert}
\newcommand{\norm}[1]{\left\lVert#1\right\rVert}

\newcommand{\setst}[2]{\left\{#1\mathrel{}\middle|\mathrel{}#2\right\}}

\newcommand{\et}{\quad \text{and} \quad}

\newcommand{\hence}{\quad \text{hence} \quad}

\newcommand{\numberset}[1]{\mathbb{#1}}

\newcommand{\Z}{\numberset{Z}}

\newcommand{\R}{\numberset{R}}
\newcommand{\C}{\numberset{C}}

\newcommand{\diff}{\mathop{}\mathopen{}\mathrm{d}}

\newcommand{\schatten}[1]{\mathfrak{S}_{#1}}
\newcommand{\Cgaugesobolev}{\dot{H}^1_\mathrm{div}\left(\R^3\right)}

\DeclareMathOperator{\tr}{tr}

\theoremstyle{plain}
\newtheorem{thm}{Theorem}[section]

\newtheorem{lem}[thm]{Lemma}

\theoremstyle{definition}

\theoremstyle{remark} 
\newtheorem{rmkx}[thm]{Remark}

\newenvironment{rmk}
  {%
   \pushQED{\qed}\begin{rmkx}}
  {\popQED\end{rmkx}}

\title[The PV regularised free energy of Dirac's vacuum in purely magnetic fields]{The Pauli-Villars regularised free energy of Dirac's vacuum in purely magnetic fields}
\author[Umberto Morellini]{Umberto Morellini}
\address{Ceremade (UMR 7534), Université Paris Dauphine - PSL, Place du Maréchal de Lattre de Tassigny, 75775 Paris Cedex 16, France}
\email{\href{mailto:morellini@ceremade.dauphine.fr}{\nolinkurl{morellini@ceremade.dauphine.fr}}}
\date{13 January 2025}

\begin{document}

\begin{abstract}
    The Dirac vacuum is a non-linear polarisable medium rather than an empty space. This non-linear behaviour starts to be significant for extremely large electromagnetic fields such as the magnetic field on the surface of certain neutron stars. Even though the null temperature case was deeply studied in the past decades, the problem at non-zero temperature needs to be better understood.\\
    In this work, we present the first rigorous derivation of the one-loop effective magnetic Lagrangian at positive temperature, a non-linear functional describing the free energy of the Dirac vacuum in a classical magnetic field. After introducing our model, we properly define the free energy functional using the Pauli-Villars regularisation technique in order to remove the worst ultraviolet divergences, which represent a well known issue of the theory. The study of the properties of this functional is addressed before focusing on the limit of slowly varying classical magnetic fields. In this regime, we prove the convergence of this functional to the Euler-Heisenberg formula with thermal corrections, recovering the effective Lagrangian first derived by Dittrich \cite{Dit-1979-PRD} in $1979$.
\end{abstract}

\maketitle

\tableofcontents

\section{Introduction}\label{sec:introduction}
The non-linear polarisable nature of the vacuum in quantum field theory has been known for a long time \cite{Dir-1934-MPCPS, EulKoc-1935-DN, Wei-1936-MFM}. In this framework, the classical action is replaced by an effective action taking the quantum corrections into account while guaranteeing the validity of the principle of least action. In quantum electrodynamics, a standard way to get the effective action is by integrating out fermions in the path integral (see for instance \cite[Ch.~33]{Sch-2013-book}). In this manner, we obtain a functional of a classical electromagnetic field treated as an external one. In \cite{GraHaiLewSer-2013-ARMA}, the authors derive in a rigorous way the effective Lagrangian action for time-independent fields in the Coulomb gauge,
\begin{multline*}
    \mathcal{L}\left(\boldsymbol{A}\right)=-\mathcal{F}_\mathrm{vac}\left(e\boldsymbol{A}\right)+e\int_{\R^3}\left(j_\mathrm{ext}\cdot A\left(x\right)-\rho_\mathrm{ext}\left(x\right)V\left(x\right)\right)\diff x\\
    +\frac{1}{8\pi}\int_{\R^3}\left(|E\left(x\right)|^2-|B\left(x\right)|^2\right)\diff x,
\end{multline*}
where $e$ is the elementary charge of an electron, $\boldsymbol{A}=\left(V,A\right)$ is a classical $\R^4-$valued electromagnetic potential with corresponding field $\boldsymbol{F}=\left(E,B\right)=\left(-\nabla V,\mathrm{curl}A\right)$, and $\rho_\mathrm{ext}$ and $j_\mathrm{ext}$ are given external charge and current densities. The corresponding Euler-Lagrange equations are the classical linear Maxwell equations with non-linear and non-local correction terms. These quantum corrections are negligible in the everyday regime and they start to be significant only for extremely large electromagnetic fields. The threshold above which the electromagnetic field starts behaving in a non-linear way is known as \textit{Schwinger limit} and it is several orders of magnitude above what can be currently produced in a laboratory ($E_c\simeq 10^{18}\,V/m$ and $B_c\simeq 10^9\,T$). To give an idea, such an electric field would accelerate a proton from rest to the maximum energy attained by protons at the Large Hydron Collider in only approximately $5$ micrometers. However, the detection of these non-linear effects is still a very active area of experimental research \cite{BurFieHorSpe-1997-PRL, MouTajBul-2006-RMP}. On the other hand, it is known that such a magnetic field strength is attained on the surface of magnetars, that is neutron stars with really large magnetic field, where these terms play an important role \cite{BarHar-2001-TAJ, DenSve-2003-AA, MarBroSte-2003-PRL}. In particular, the observation of one of these effects has been recently announced \cite{FanKamInaYam-2017-EPJD, MigTestGonTav-2016-MNRAS}.\par
The vacuum energy $\mathcal{F}_\mathrm{vac}$ is a complicated non-local functional which suffers from ultraviolet divergences. In \cite{GraHaiLewSer-2013-ARMA}, the authors deal with them by using the Pauli-Villars regularisation method in order to give a rigorous definition of the functional. A simple and useful approximation, very well known in the physics literature, consists in replacing the complicated vacuum energy term $\mathcal{F}_\mathrm{vac}$ by a superposition of local independent problems, that is
\begin{equation}\label{eq:superposition-vacuum-energy}
    \mathcal{F}_\mathrm{vac}\simeq\int_{\R^3}f_\mathrm{vac}\left(eE\left(x\right),eB\left(x\right)\right)\diff x,
\end{equation}
where $f_\mathrm{vac}\left(eE,eB\right)$ is the energy per unit volume for a constant electromagnetic field $\boldsymbol{F}=\left(E,B\right)$. The function $f_\mathrm{vac}$ was first computed by Euler and Heisenberg \cite{HeiEul-1936-ZFP} in $1936$. Some alternative derivations were then presented by  Weisskopf \cite{Wei-1936-MFM} and Schwinger \cite{Sch-1951-PR}, who introduced a largely renowned method in the physics literature known as \textit{proper time method}. Since the integrand in the Euler-Heisenberg formula \cite[Formula~4]{GraLewSer-2018-JMPA} can have poles on the real line, a proper definition requires to shift the integration path to the complex plane by replacing the integral variable $s$ by $s+i\eta$ and taking the limit $\eta\xrightarrow{}0$ \cite{Sch-1951-PR}. In this manner, the function $f_\mathrm{vac}$ gains an exponentially small non-zero imaginary part which was interpreted by Schwinger as the electron-positron pair production rate and explains the instability of the vacuum \cite{DitGie-2000-book}\cite[Paragraph~7.3]{GreRei-2009-book}. Nevertheless, in the case of a purely magnetic field, the Euler-Heisenberg formula is absolutely convergent and real, as expected. For this reason, this situation has been deeply studied in the physics literature \cite{Con-1972-NPB, TsaErb-1974-PRD, TsaErb-1975-PRD, MelSto-1976-INCA} and the function $f_\mathrm{vac}$ can be simplified as follows:
\begin{equation}\label{eq:physics-magnetic-EH}
    f_\mathrm{vac}\left(0,eB\right)=\frac{1}{8\pi^2}\int_0^\infty \frac{e^{-sm^2}}{s^3}\left(es|B|\coth\left(es|B|\right)-1-\frac{e^2 s^2 |B|^2}{3}\right)\diff s.
\end{equation}\par
In the non-perturbative physical examples above, the fields exist in a thermal bath or in some non-equilibrium background which is very different from the vacuum. For this reason, the study of finite temperature effects in quantum field theory attracted the interest of many physicists in the past decades. In particular, several authors \cite{KirLin-1972-PLB, Wei-1974-PRD, DolJac-1974-PRD} considered what happens when a system of elementary particles described by a quantum field theory is heated. They found that symmetries which are spontaneously broken at zero temperature (such as those of the weak interactions) may be restored at sufficiently high temperatures, and calculated the critical temperature at which such a restoration takes place. To do this kind of calculation, one needs to know the Feynman rules for a field theory at finite temperature. For a non-gauge theory, these rules can be derived using well known methods \cite{FetWal-1971-book}. However, for a gauge theory, a more powerful technique is needed to cope with several new problems that arise \cite{Ber-1974-PRD}. In any case, the Feynman rules at finite temperature $T$ are strictly related to the Kubo-Martin-Schwinger (KMS) condition (see \cref{stat:KMS-condition}) and consist in the zero temperature rules with the following formal replacements:
\begin{align}\notag
    \int\frac{\diff^4 p}{\left(2\pi\right)^4}&=\frac{i}{\beta}\sum_n\int\frac{\diff^3 p}{\left(2\pi\right)^3},\\\label{eq:T-Feynman-rules}
    p_0&=i\omega_n,\\\notag
    \left(2\pi\right)^4\delta^4\left(p_1+p_2+\ldots\right)&=\frac{\beta}{i}\left(2\pi\right)^3\delta\left(\omega_{n_1}+\omega_{n_2}+\ldots\right)\delta^3\left(\Vec{p_1}+\Vec{p_2}+\ldots\right),
\end{align}
where $\beta=1/T$ and
\[
\omega_n=
\begin{cases}
    \frac{2n\pi i}{\beta}\quad\text{for bosons},\\
    \frac{\left(2n+1\right)\pi i}{\beta}\quad\text{for fermions}.
\end{cases}
\]
In $1979$, Dittrich studied thermal effects in quantum electrodynamics. In particular, in \cite{Dit-1979-PRD}, he calculates the one-loop effective potential at finite temperature for scalar and spinor QED in the presence of a constant magnetic field, namely the Euler-Heisenberg Lagrangian at positive temperature. In his work, he shows that the one-loop effective potential at finite temperature represents - in the language of thermodynamics - the contribution of the vacuum energy to the total free energy in presence of an external constant field. By employing the proper time method and using the replacements \eqref{eq:T-Feynman-rules}, the following Lagrangian at finite temperature is derived:
\[
\mathcal{L}\left(B,T\right)=\frac{\left(\pi i\right)^{1/2}}{4\pi^2\beta}\int_0^\infty e^{-ism^2} \left(esB\right)\cot\left(esB\right)\sum_{n=-\infty}^\infty e^{is\omega_n^2}\frac{\diff s}{s^{5/2}},
\]
which is obviously an ill-defined function. Notice that the cotangent here can be explained by a simple rotation of the integral domain in the imaginary plane as it will be discussed later. This is a standard technique in the physics literature allowing to avoid the poles of the integrand function. This explains also why the imaginary exponential appears in the integral above. After some further manipulations, this formula can be expressed as
\begin{equation}\label{eq:lagrangian-decomposition-thermal-contribution}
    \mathcal{L}\left(B,T\right)=\mathcal{L}^0\left(B\right)+\mathcal{L}^T\left(B,T\right),
\end{equation}
where $\mathcal{L}^0$ is the Euler-Heisenberg Lagrangian at null temperature given by \eqref{eq:physics-magnetic-EH} and the term $\mathcal{L}^T$, containing all the thermal effects, is expressed as follows:
\begin{equation}\label{eq:T-physics-magnetic-EH}
    \mathcal{L}^T\left(B,T\right)=\frac{1}{8\pi^2}\int_0^\infty e^{-ism^2}\left(esB\right)\cot\left(esB\right)\sum_{n\neq0}\left(-1\right)^n e^{i\frac{n^2\beta^2}{4s}}\frac{\diff s}{s^3}.
\end{equation}
This result can be related to well known findings of statistical thermodynamics, by studying the case of massless fermions (and bosons). In order to do so, setting $B=0$ and $m=0$ in $\mathcal{L}^T$ yields
\[
\mathcal{L}_F\left(0,T\right)=\frac{2}{3}\frac{7}{120}\pi^2k^4T^4,
\]
where $\mathcal{L}_F$ stands for the fermionic choice of $\omega_n$ and $k$ is the Boltzmann constant. Thanks to the following formula:
\[
\int_0^\infty\frac{x^{2n-1}}{e^x+1}\diff x=\left(1-2^{1-2n}\right)\left(2\pi\right)^{2n}\frac{\abs{B_{2n}}}{4n},
\]
where $B_n$ are the Bernoulli numbers, one can finally gets
\[
\mathcal{L}_F\left(0,T\right)=\frac{2}{3}\frac{1}{\pi^2}\int_0^\infty \frac{x^3}{e^{x/kT}+1}\diff x,
\]
which is the correct Fermi-Dirac distribution, confirming the coherence of these calculations within the statistical framework. Later, the thermal effective action for a constant magnetic field  \cite{Dit-1979-PRD, ElmPerSka-1993-PRL} has been generalised to arbitrary constant electromagnetic fields in \cite{ElmSka-1995-PLB}. In other words, here the one-loop effective action for a constant electromagnetic field \cite{Sch-1951-PR} has been generalised to finite temperature and chemical potential.\par
It is now clear that the calculation of the effective action at positive temperature has been an interesting physical question for some decades, particularly because the imaginary part of the effective action is expected to give the vacuum decay rate and one would like to understand the influence of the temperature on this rate. There have been several attempts to generalise Schwinger's proper time method to the finite temperature case in order to determine the temperature dependent effective actions, leading to conflicting results in the physics literature. The subtleties behind these results have to do with the imposition of the anti-periodicity condition required at positive temperature as well as on the finite temperature formalism used to carry out the calculation. We refer the interested reader to \cite{Das-2011-JPCS}, based on the previous papers \cite{DasFre-2009-PRD, DasFre-2010-PRD, DasFre-2011-PLB}, for more details about the argument. We also refer the reader to \cite[Chapter~3.5]{DitGie-2000-book} for a complete review about QED at finite temperature.\par
The aim of this paper is first of all to properly define the vacuum free energy $\mathcal{F}_\mathrm{vac}^T$ of a quantised Dirac field at positive temperature $T$. Again, this quantity suffers from divergences in the high energy regime which need to be regularised. In order to do so, we use the Pauli-Villars technique as in \cite{GraHaiLewSer-2013-ARMA}.\\
In \cref{sec:derivation-of-the-model-and-main-results}, we derive the model starting from the Hamiltonian of a second-quantised fermionic field in a given electromagnetic potential $\boldsymbol{A}=\left(V,A\right)$. After obtaining the vacuum energy, we then consider the associated vacuum free energy by introducing an entropy term. The whole framework is discussed in detail and main results are presented.\\
In \cref{sec:rigorous-definition-of-the-PV-regularised-free-energy-functional}, the first part of \cref{stat:main-thm-1} stating the well-posedness of the vacuum free energy functional is proved. Then, thanks to additional gauge-invariant estimates, we give a proof of the second part of the theorem providing the main properties of the addressed functional and allowing us to extend it to the most natural function space for this problem. In particular, we find replacements similar to \eqref{eq:T-Feynman-rules} which need to be applied to the null temperature setting \cite{GraHaiLewSer-2013-ARMA, GraLewSer-2018-JMPA} in order to get the right calculations for the non-zero temperature case. Due to these prescriptions, the integration by parts with respect to the discretised variable does not hold anymore adding some additional technical difficulties to this problem. Partly for this reason, we restrict our study to the purely magnetic case. Moreover, \cref{stat:explicit-second-order} shows that we cannot expect an effect similar to the one shown in \cite{HaiLewSei-2008-RMP} for the reduced Bogoliubov-Dirac-Fock (BDF) model, where the authors prove that the thermal effects make the particles rearrange in the polarised Dirac sea in order to completely screen the external potential. This is a classical phenomenon in the field of non-relativistic fermionic plasma physics, known as \textit{Debye screening}.\\
Later, in \cref{sec:the-regime-of-slowly-varying-magnetic-fields}, we provide the first rigorous derivation of the purely magnetic Euler-Heisenberg formula for the vacuum free energy at positive temperature, in the regime where $B$ is slowly varying in space. In this limit, we prove the convergence of the free energy to the null temperature Euler-Heisenberg formula corrected by thermal contributions, recovering a Pauli-Villars regularised effective Lagrangian which coincides with the one first derived by Dittrich \cite{Dit-1979-PRD} in $1979$ (see \eqref{eq:lagrangian-decomposition-thermal-contribution} and \eqref{eq:T-physics-magnetic-EH}), up to some standard mathematical manipulations needed so as to give a rigorous meaning to this object, as it will be discussed in this section. Even though the null temperature case has been rigorously studied in \cite{GraLewSer-2018-JMPA}, to the best of our knowledge the non-zero temperature problem has been treated only in the physics literature. More specifically, we consider a magnetic field of the form $B\left(\varepsilon x\right)$ and look at the limit $\varepsilon\xrightarrow{}0$. Notice that in this context we get a strong magnetic potential given by $A_\varepsilon\left(x\right)=\varepsilon^{-1}A\left(\varepsilon x\right)$. Nevertheless, this is not a source of additional difficulties from a spectral point of view (see \cref{stat:Dirac-spectral-properties}). Indeed, briefly speaking, the presence of a purely magnetic field ensures us to keep the spectrum of the Dirac operator far apart from zero without adding any additional hypothesis on the size of the potential (as we would expect to do if an electric potential was turned on). In this regime, as stated in \cref{stat:main-thm-2} below, we recover an asymptotic Euler-Heisenberg formula for the vacuum free energy of the following form:
\[
\mathcal{F}_\mathrm{vac}^T\left(0,eA_\varepsilon\right)\underset{\varepsilon\xrightarrow{}0}{\sim}\int_{\R^3}f_\mathrm{vac}^T\left(0,eB\left(\varepsilon x\right)\right)\diff x=\varepsilon^{-3}\int_{\R^3}f_\mathrm{vac}^T\left(0,eB\left(x\right)\right)\diff x,
\]
where $f^T_\mathrm{vac}$ is a suitable function describing the free energy per unit volume.

\section{Derivation of the model and main results}\label{sec:derivation-of-the-model-and-main-results}
We want to define the free energy of QED vacuum at positive temperature. In order to do so, we first consider a fermionic second-quantised field placed in a given electromagnetic potential $\boldsymbol{A}=\left(V,A\right)$. The potential is kept fixed and we look for the ground state energy of the Dirac field in the given $\boldsymbol{A}$. The Hamiltonian of the field reads
\begin{equation}\label{eq:field-hamiltonian}
    \mathbb{H}^{e\boldsymbol{A}}=\frac{1}{2}\int_{\R^3}\left(\Psi^*\left(x\right)D^{e\boldsymbol{A}}_m\Psi\left(x\right)-\Psi\left(x\right)D^{e\boldsymbol{A}}_m\Psi^*\left(x\right)\right)\diff x,
\end{equation}
where $\Psi\left(x\right)$ is the second-quantised field operator which can physically be interpreted as the annihilation of an electron at $x$ and mathematically is defined by means of the following anticommutation relation:
\[
\Psi^*\left(x\right)_\sigma \Psi\left(y\right)_\nu +\Psi\left(y\right)_\nu \Psi^*\left(x\right)_\sigma =2\delta_{\sigma,\nu}\delta\left(x-y\right).
\]
Here $1\leq\sigma,\nu\leq 4$ are the spin variables and $\Psi\left(x\right)_\sigma$ is an operator-valued distribution. The Hamiltonian $\mathbb{H}^{e\boldsymbol{A}}$ formally acts on the fermionic Fock space,
\[
\mathcal{F}=\C\oplus\bigoplus_{N\geq 1}\bigwedge_1^N L^2\left(\R^3,\C^4\right),
\]
and has the interesting property of being charge conjugation invariant,
\[
C\mathbb{H}^{e\boldsymbol{A}}C^{-1}=\mathbb{H}^{-e\boldsymbol{A}}
\]
(where $C$ is the charge-conjugation operator in Fock space as defined in \cite[Formula~(1.81)]{Tha-1992-book}), when the integrand in \eqref{eq:field-hamiltonian} is interpreted as follows:
\begin{multline*}
    \Psi^*\left(x\right)D^{e\boldsymbol{A}}_m\Psi\left(x\right)-\Psi\left(x\right)D^{e\boldsymbol{A}}_m\Psi^*\left(x\right)\\
    =\sum_{\mu,\nu=1}^4\left(\Psi^*\left(x\right)_\mu \left(D^{e\boldsymbol{A}}_m\right)_{\mu,\nu}\Psi\left(x\right)_\nu-\Psi\left(x\right)_\mu\left(D^{e\boldsymbol{A}}_m\right)_{\mu,\nu}\Psi^*\left(x\right)_\nu\right).
\end{multline*}\par
In formula \eqref{eq:field-hamiltonian},
\[
D^{e\boldsymbol{A}}_m=\boldsymbol{\alpha}\cdot\left(-i\nabla-eA\right)+eV+m\beta
\]
is the electromagnetic Dirac operator acting on $L^2\left(\R^3,\C^4\right)$. We work in a system of units such that the speed of light and Planck's reduced constant are both set equal to one, $c=\hbar=1$. The four Dirac matrices $\boldsymbol{\alpha}=\left(\alpha_1,\alpha_2,\alpha_3\right)$ and $\beta$ are given by
\[
\alpha_k=\begin{bmatrix}
0 & \sigma_k\\
\sigma_k & 0
\end{bmatrix},\quad
\beta=\begin{bmatrix}
I_2 & 0\\
0 & -I_2
\end{bmatrix},
\]
with Pauli matrices $\sigma_1$, $\sigma_2$ and $\sigma_3$ defined as
\[
\sigma_1=\begin{bmatrix}
0 & 1\\
1 & 0
\end{bmatrix},\quad
\sigma_2=\begin{bmatrix}
0 & -i\\
i & 0
\end{bmatrix},\quad
\sigma_3=\begin{bmatrix}
1 & 0\\
0 & -1
\end{bmatrix}.
\]
The spectrum of the free Dirac operator is not bounded from below \cite{Tha-1992-book}. Indeed,
\[
\sigma\left(D^0_m\right)=\intoc{-\infty}{-m}\cup\intco{m}{\infty}.
\]
This led Dirac to postulate that the vacuum is filled with infinitely many virtual particles occupying the negative energy states (\textit{Dirac sea}) \cite{Dir-1930-PRSL, Dir-1934-MPCPS, Dir-1934-SR}. As a consequence, a real free electron cannot be in a negative state due to the Pauli exclusion principle. Moreover, Dirac conjectured the existence of "holes" in the Dirac sea interpreted as positrons, having a positive charge and a positive energy. Dirac also predicted the phenomenon of vacuum polarisation: in the case of an external electric field, the virtual electrons align their charges with the field direction and the vacuum obtains a non-constant density of charge. Actually the polarised vacuum modifies the electrostatic field and the virtual electrons react to the corrected field. We refer the reader to \cite{EstLewSer-2008-BAMS} for a detailed review about the difficulties arising from the negative part of the spectrum of the free Dirac operator.\par
Returning to the electromagnetic Dirac operator, it is natural then to introduce the following Coulomb-gauge homogeneous Sobolev space:
\[
\Cgaugesobolev=\setst{\boldsymbol{A}=\left(V,A\right)\in L^6\left(\R^3,\R^4\right)}{\mathrm{div}A=0\text{ and }\boldsymbol{F}=\left(-\nabla V,\mathrm{curl}A\right)\in L^2\left(\R^3,\R^6\right)},
\]
endowed with its norm
\[
\|\boldsymbol{A}\|^2_{\Cgaugesobolev}=\|\nabla V\|^2_{L^2\left(\R^3\right)}+\|\mathrm{curl}A\|^2_{L^2\left(\R^3\right)}=\|\boldsymbol{F}\|^2_{L^2\left(\R^3\right)}.
\]
The equation $\mathrm{div}A=0$ should be interpreted in a distributional sense. The assumption on $\boldsymbol{F}$ being in $L^2\left(\R^3\right)$ trivially means that the electromagnetic field has a finite energy,
\[
\int_{\R^3}|E|^2+|B|^2 <\infty.
\]
The proof of the following lemma, which recalls some spectral properties of the electromagnetic Dirac operator $D^{e\boldsymbol{A}}_m$, can be found in \cite[Lemma~2.1]{GraHaiLewSer-2013-ARMA}:
\begin{lem}\label{stat:Dirac-spectral-properties}
    Let $m>0$ and $\boldsymbol{A}\in\Cgaugesobolev$.
    \begin{itemize}
        \item[(i)] The operator $D^{e\boldsymbol{A}}_m$ is self-adjoint on $H^1\left(\R^3,\C^4\right)$ and its essential spectrum is
        \[
        \sigma_\mathrm{ess}\left(D^{e\boldsymbol{A}}_m\right)=\intoc{-\infty}{-m}\cup\intco{m}{\infty}.
        \]
        \item[(ii)] There exists a universal constant $C$ such that, if $\|\boldsymbol{A}\|_{\Cgaugesobolev}\leq\eta\sqrt{m}$, for some number $\eta<1/C$, then
        \[
        \sigma\left(D^{e\boldsymbol{A}}_m\right)\cap\intoo{-m\left(1-C\eta\right)}{\left(1-C\eta\right)m}=\varnothing.
        \]
        \item[(iii)] Finally, if $V=0$, then $\sigma\left(D^{e\boldsymbol{A}}_m\right)\cap\intoo{-m}{m}=\varnothing$.
    \end{itemize}
\end{lem}
The expectation value of the energy in any state in Fock space can be expressed as \cite[Equation~(2.8)]{GraHaiLewSer-2013-ARMA}
\begin{equation}\label{eq:expectation-value-energy}
    \langle\mathbb{H}^{e\boldsymbol{A}}\rangle=\tr_{L^2\left(\R^3,\C^4\right)}\left[D^{e\boldsymbol{A}}_m\left(\gamma-\frac{1}{2}\right)\right]=E\left(\gamma\right),
\end{equation}
where $\gamma$ is the one-particle density matrix associated to a given state, namely
\[
\gamma\left(x,y\right)_{\sigma,\nu}=\langle\Psi^*\left(x\right)_\sigma \Psi\left(y\right)_\nu\rangle.
\]
The renormalised density matrix in \eqref{eq:expectation-value-energy} is a consequence of charge-conjugation invariance. The interested reader can find more details in \cite{HaiLewSol-2007-CPAM}. Since electrons are fermions, they obey the Pauli exclusion principle, which implies that $\gamma$ has to satisfy $0\leq\gamma\leq 1$ on $L^2\left(\R^3,\C^4\right)$. On the other hand, any operator $\gamma$ verifying $0\leq\gamma\leq 1$ formally arises from one state in Fock space. Since the energy depends on the state of the electrons only through the density matrix $\gamma$, we can reformulate our problem by focusing on the simpler operator $\gamma$ and the corresponding energy \eqref{eq:expectation-value-energy}.\par
Recall that we are interested in finding the ground state of the vacuum at positive temperature $T$, therefore we need to consider the following free energy:
\begin{equation}\label{eq:definition-free-energy}
    F\left(\gamma\right)=E\left(\gamma\right)-T \tr\left[S\left(\gamma\right)\right],
\end{equation}
where
\[
S\left(x\right)=-x\log x-\left(1-x\right)\log\left(1-x\right)
\]
and therefore $\tr\left[S\left(\gamma\right)\right]$ is an entropy term which is well known in the literature (see e.g. \cite[Formula~(2c.10)]{BacLieSol-1994-JSP}, \cite[Formula~(4)]{HaiLewSei-2008-RMP} and \cite[Formula~(5)]{GonKouSer-2023-AHP}). Then, the minimisation of the free energy functional with respect to $\gamma$ has to be addressed. For atoms and molecules, one should add a charge constraint of the following form:
\[
\tr\left(\gamma-\frac{1}{2}\right)=N.
\]
Nevertheless, in this work, we restrict ourselves to the vacuum case for simplicity. Therefore, we do not have any other constraint than $0\leq\gamma\leq 1$. In this section, we set the electric potential $V$ to be equal to $0$ and focus on the derivation of Dirac's vacuum free energy in an external magnetic field at positive temperature $T$. The reason of this choice will be clearer in the following sections, but essentially working with a magnetic Dirac operator ensures us to keep the spectrum far away from $0$ without adding some additional hypothesis on the size of the potential and allowing us to localise the energy to sets of fixed size. By \cite[Lemma~1.1]{GonKouSer-2023-AHP}, we know that the only critical point $\gamma^*$ of functional \eqref{eq:definition-free-energy}, which is also the unique minimiser by convexity, is given by
\[
\gamma^*=\frac{1}{1+e^{D^{eA}_m /T}}=\frac{e^{-D^{eA}_m /2T}}{2\cosh\left(D^{eA}_m /2T\right)},\hence 1-\gamma^*=\frac{1}{1+e^{-D^{eA}_m /T}}=\frac{e^{D^{eA}_m /2T}}{2\cosh\left(D^{eA}_m /2T\right)}.
\]
Then, the corresponding value of the free energy formally is
\begin{align}\notag
    F\left(\gamma^*\right)&=\tr\left[D^{eA}_m\left(\gamma^*-\frac{1}{2}\right)\right]-T\tr\left[S\left(\gamma^*\right)\right]\\ \label{eq:infinite-free-energy}
    &=-\frac{1}{\beta}\tr\left[\log\left(2\cosh\frac{\beta D^{eA}_m}{2}\right)\right]=\mathcal{F}\left(eA,\beta\right).
\end{align}
Of course this free energy is infinite except if our model is settled in a box with an ultraviolet cut-off \cite{HaiLewSol-2007-CPAM}. In order to give a clear mathematical meaning to \eqref{eq:infinite-free-energy}, we proceed as follows. First, we can subtract the (infinite) free energy of the free Dirac sea and define the relative free energy as
\begin{equation}\label{eq:relative-free-energy}
    \mathcal{F}_\mathrm{rel}\left(eA,\beta\right)=\frac{1}{\beta}\tr\left[\log\left(2\cosh\frac{\beta D^{0}_m}{2}\right)-\log\left(2\cosh\frac{\beta D^{eA}_m}{2}\right)\right].
\end{equation}
By subtracting an infinite constant, we do not formally change the variational problem that we are interested in, so we also do not change the resulting equations. However, the functional \eqref{eq:relative-free-energy} is not well defined yet because of the well known ultraviolet divergences of this theory. Indeed, the operator $\log\left(2\cosh\left(\beta D^{0}_m/2\right)\right)-\log\left(2\cosh\left(\beta D^{eA}_m/2\right)\right)$ is not trace class as long as $A\neq 0$. This can formally be seen by expanding the trace in a power series of $eA$. As we will see later, the first order term vanishes and the second order term is infinite due to ultraviolet divergences. Moreover, the formal higher order terms cannot prevent this ultraviolet divergence.\par
It is then clear that a UV cut-off has to be imposed. The choice of this regularisation is extremely important. As shown below, in the expansion of $\eqref{eq:relative-free-energy}$ as a power series of $eA$, several terms vanish due to gauge invariance. This is an extra reason to preserve the gauge symmetry, in addition to well known physical motivations. In \cite{GraLewSer-2009-CMP}, the authors studied two ways of dealing with UV divergences in the purely electrostatic case, but both of them would not work here because of their lack of gauge symmetry.\par
Here, we will use the regularisation technique introduced by Pauli and Villars \cite{PauVil-1949-RMP} in $1949$. Of course, this is not the only possible choice when gauge invariance has to be conserved (see for instance \cite{Lei-1975-RMP} for an alternative approach). The Pauli-Villars method consists in introducing $J$ fictitious particles into the model with very high masses $m_1,\ldots,m_J$ playing the role of ultraviolet cut-offs. Indeed, note that $m$ has the dimension of a momentum since in our system of units $\hbar=c=1$. The only role of these additional particles is to regularise the model at high energies and they have no physical meaning. Indeed, these particles being really massive, they do not affect the low energy regime. In our framework, the Pauli-Villars method consists in introducing the following energy functional:
\begin{equation}\label{eq:PV-free-energy}
    \mathcal{F}_\mathrm{PV}\left(eA,\beta\right)=\frac{1}{\beta}\tr\left[\sum_{j=0}^J c_j\left(\log\left(2\cosh\frac{\beta D^{0}_{m_j}}{2}\right)-\log\left(2\cosh\frac{\beta D^{eA}_{m_j}}{2}\right)\right)\right].
\end{equation}
Here, $m_0=m$ and $c_0=1$, and the coefficients $c_j$ and $m_j$ are chosen such that
\begin{equation}\label{eq:PV-conditions}
    \sum_{j=0}^J c_j=\sum_{j=0}^J c_j m_j^2=0.
\end{equation}
It is well known in the physics literature \cite{PauVil-1949-RMP,GreRei-2009-book,BjoDre-1965-book} that only two auxiliary fields are necessary to fulfill these conditions, hence we shall take $J=2$ in the rest of this work. In this case, condition \eqref{eq:PV-conditions} is equivalent to
\[
c_1=\frac{m_0^2-m_2^2}{m_2^2-m_1^2}\et c_2=\frac{m_1^2-m_0^2}{m_2^2-m_1^2}.
\]
We shall always assume that $m_0<m_1<m_2$, which implies that $c_1<0$ and $c_2>0$. The role of the constraint \eqref{eq:PV-conditions} is to remove the worst linear ultraviolet divergences. Indeed, the Pauli-Villars regularisation does not avoid a logarithmic divergence when $m_1,m_2\xrightarrow{}\infty$. In order to better understand this, we can define the averaged ultraviolet cut-off $\Lambda$ as
\begin{equation}\label{eq:averaged-UV-cutoff}
    \log\left(\Lambda^2\right)=-\sum_{j=0}^2 c_j\log\left(m_j^2\right).
\end{equation}
Of course, once the value of $\Lambda$ is fixed, we cannot uniquely determine $m_1,m_2$. Practically speaking, we usually choose the masses as functions of $\Lambda$ such that the Pauli-Villars coefficients $c_1,c_2$ stay bounded when $\Lambda\xrightarrow{}\infty$. As already mentioned, the logarithmic divergence in the averaged cut-off $\Lambda$ can explicitly be seen in the second order term in the expansion which will be studied later in this work (see \cref{subsec:gauge-invariant-estimates}).\par
Now, in order to simplify our calculations, we rewrite the Pauli-Villars-regularised free energy functional in an integral form. In order to do so, we derive \eqref{eq:PV-free-energy} with respect to $\beta$ by getting:
\[
\frac{\partial}{\partial\beta}\left(\beta \mathcal{F}_\mathrm{PV}\left(eA,\beta\right)\right)=\frac{1}{2}\tr\left[\sum_{j=0}^2 c_j \left(D^{0}_{m_j} \tanh\frac{\beta D^{0}_{m_j}}{2}-D^{eA}_{m_j} \tanh\frac{\beta D^{eA}_{m_j}}{2}\right)\right],
\]
which implies
\begin{equation}\label{eq:integral-free-energy}
    \mathcal{F}_\mathrm{PV}\left(eA,\beta\right)=\frac{1}{2\beta}\int_0^\beta \tr\left[\sum_{j=0}^2 c_j \left(D^{0}_{m_j} \tanh\frac{b D^{0}_{m_j}}{2}-D^{eA}_{m_j} \tanh\frac{b D^{eA}_{m_j}}{2}\right)\right]\diff b.
\end{equation}
Notice that thanks to conditions \eqref{eq:PV-conditions} the value of $\mathcal{F}_\mathrm{PV}$ goes to zero as $\beta$ goes to zero. Let us remark that in the following the elementary charge $e$ will be set to be equal to $1$.\par
We can now state one of the main results of this paper stating the well-posedness of the PV-regularised free energy functional $\mathcal{F}_\mathrm{PV}$ and giving its properties. The proof of this theorem will be given in \cref{sec:rigorous-definition-of-the-PV-regularised-free-energy-functional}.
\begin{thm}\label{stat:main-thm-1}
    Assume that $c_j$ and $m_j$ satisfy
    \[
    c_0=1,\quad m_2>m_1>m_0>0 \et \sum_{j=0}^2 c_j=\sum_{j=0}^2 c_j m_j^2=0.
    \]
    Let
    \[
    T_A\left(\beta\right)=\frac{1}{2}\sum_{j=0}^2 c_j\left(D^0_{m_j}\tanh{\frac{\beta D^0_{m_j}}{2}}-D^A_{m_j}\tanh{\frac{\beta D^A_{m_j}}{2}}\right).
    \]
    \begin{itemize}
        \item[(i)] For any $A\in L^1\left(\R^3,\R^3\right)\cap H^1\left(\R^3,\R^3\right)$, the operator $T_A\left(\beta\right)$ is trace class on $L^2\left(\R^3,\C^4\right)$, well defined with respect to $\beta$ and bounded in a neighbourhood of $\beta=0$. In particular, $\mathcal{F}_\mathrm{PV}\left(A,\beta\right)$ is well defined in this case by
        \[
        \mathcal{F}_\mathrm{PV}\left(A,\beta\right)=\frac{1}{\beta}\int_0^\beta\tr\left[T_A\left(b\right)\right]\diff b.
        \]
        \item[(ii)] Let $A\in L^1\left(\R^3,\R^3\right)\cap\Cgaugesobolev$. We have
        \[
        \mathcal{F}_\mathrm{PV}\left(A,\beta\right)=\mathcal{F}_{\mathrm{PV},2}\left(B,\beta\right)+\mathcal{R}\left(A,\beta\right),
        \]
        where $B=\mathrm{curl}A$. The functional $\mathcal{R}$ is of class $C^\infty$ on $\Cgaugesobolev$ and there exists a universal constant $\kappa=\kappa\left(\beta\right)$ such that
        \begin{equation}\label{eq:remainder-estimate}
            \abs{\mathcal{R}\left(A,\beta\right)}\leq \kappa\left(\left(\sum_{j=0}^2\frac{\abs{c_j}}{m_j}\right)\norm{B}^4_{L^2\left(\R^3\right)}+\left(\sum_{j=0}^2\frac{\abs{c_j}}{m_j^2}\right)\norm{B}^6_{L^2\left(\R^3\right)}\right).
        \end{equation}
        The functional $\mathcal{F}_{\mathrm{PV},2}$ is the nonnegative and bounded quadratic form on $L^2\left(\R^3,\R^3\right)$ given by
        \[
        \mathcal{F}_{\mathrm{PV},2}\left(B,\beta\right)=\frac{1}{8\pi}\int_{\R^3}\left(M^0\left(q\right)+M^T\left(q,\beta\right)\right)\abs{\widehat{B}\left(q\right)}^2\diff q,
        \]
        with
        \[
        M^0\left(q\right)=-\frac{2}{\pi}\int_0^1\sum_{j=0}^2 c_j\log\left(m_j^2+u\left(1-u\right)q^2\right)u\left(1-u\right)\diff u
        \]
        and
        \begin{multline*}
            M^T\left(q,\beta\right)=-\frac{8}{\pi}\int_0^1\int_0^\infty\sum_{j=0}^2 c_j\left[\frac{1}{1+e^{-X_j\left(\beta,u,q\right)\cosh t}}+\frac{1}{X_j\left(\beta,u,q\right)\cosh t}\left(2\log 2\right.\right.\\
            \left.\left.-\log\left(\left(e^{X_j\left(\beta,u,q\right)\cosh t}+1\right)\left(e^{-X_j\left(\beta,u,q\right)\cosh t}+1\right)\right)\right)\right]\diff t\,u\left(1-u\right)\diff u,
        \end{multline*}
        where $X_j\left(b,u,q\right)=b\sqrt{m_j^2+u\left(1-u\right)q^2}$ for $j=0,1,2$.\\
        Indeed, the function $M^0$ is well defined and positive on $\R^3$, and satisfies
        \begin{equation}\label{eq:M0-estimates}
            0<M^0\left(q\right)\leq M^0\left(0\right)=\frac{2\log\Lambda}{3\pi},
        \end{equation}
        where $\Lambda$ is defined by \eqref{eq:averaged-UV-cutoff}. Moreover, $M^T$ is well defined, bounded for any fixed $\beta\in\intoo{0}{+\infty}$ and such that
        \begin{equation}\label{eq:M0-MT-estimate}
            M^0\left(q\right)+M^T\left(q,\beta\right)\geq 0.
        \end{equation}
        In particular, the functional $\mathcal{F}_\mathrm{PV}$ can be uniquely extended to a continuous mapping on $\dot{H}^1_\mathrm{div}\left(\R^3\right)$.
    \end{itemize}
\end{thm}
The function $M^0$ describes the linear response of the Dirac sea at zero temperature and is very well known in the physics literature \cite[Eq.~(5.39)]{GreRei-2009-book}. From a mathematical point of view, it has been proved in \cite{GraHaiLewSer-2013-ARMA} that
\[
\underset{\Lambda\xrightarrow{}\infty}{\lim}\left(\frac{2\log\Lambda}{3\pi}-M^0\left(q\right)\right)=U\left(q\right)\coloneqq\frac{\abs{q}^2}{4\pi}\int_0^1\frac{z^2-z^4/3}{1+\abs{q}^2\left(1-z^2\right)/4}\diff z,
\]
where the function $U$ in the right-hand side was first computed by Serber \cite{Ser-1935-PR} and Uehling \cite{Ueh-1935-PR}. The same function appears in previous mathematical works studying the purely electrostatic case \cite{HaiSie-2003-CMP, HaiLewSer-2005-JPA, GraLewSer-2011-CMP}. As already mentioned, the Pauli-Villars regularisation allows us to remove only the worst linear ultraviolet divergences. Indeed, the second order term is still logarithmically divergent in the high frequencies domain because of $M^0$, as shown in \eqref{eq:M0-estimates}. On the other hand, $M^T$ is the contribution at finite temperature $T$ to the linear response of Dirac's vacuum. Estimate \eqref{eq:M0-MT-estimate} guarantees $\mathcal{F}_\mathrm{PV}$ to be a well defined free energy functional from both a mathematical and a physical point of view. For all of these reasons, \cref{stat:main-thm-1} is the equivalent of \cite[Theorem~2.1]{GraHaiLewSer-2013-ARMA} at positive temperature $T$ and will be discussed much more in detail in \cref{sec:rigorous-definition-of-the-PV-regularised-free-energy-functional}.\par
The boundedness of the second order term $\mathcal{F}_{\mathrm{PV},2}$ and of the remainder $\mathcal{R}$ allows us to uniquely extend the PV-regularised free energy functional to the whole space $\Cgaugesobolev$. This property is important for the second main result of this work, which derives an Euler-Heisenberg formula for the vacuum free energy starting from the PV-regularised free energy functional, in the regime of slowly varying purely magnetic fields. Thus, we consider a scaled magnetic field of the form $B_\varepsilon\left(x\right)=B\left(\varepsilon x\right)$, with a given $B=\mathrm{curl}A\in L^2\left(\R^3\right)$ and $A\in\Cgaugesobolev$. The following theorem requires a bit more regularity:
\begin{thm}\label{stat:main-thm-2}
    Let $B\in C^0\left(\R^3,\R^3\right)$ be such that $\mathrm{div}B=0$,
    \begin{equation}\label{eq:main-thm-2-hp}
        B\in L^1\left(\R^3\right)\cap L^\infty\left(\R^3\right),\quad\nabla B\in L^1\left(\R^3\right)\cap L^6\left(\R^3\right),
    \end{equation}
    and let $A$ be the associated magnetic potential in $\Cgaugesobolev$. Let $A_\varepsilon\left(x\right)=\varepsilon^{-1}A\left(\varepsilon x\right)$. Then,
    \[
    \varepsilon^3\mathcal{F}_\mathrm{PV}\left(A_\varepsilon,\beta\right)=\int_{\R^3}\left(f_\mathrm{PV}^0\left(\abs{B\left(x\right)}\right)+f_\mathrm{PV}^T\left(\abs{B\left(x\right)},\beta\right)\right)\diff x+\bigO{\varepsilon},
    \]
    where
    \[
    f_\mathrm{PV}^0\left(a\right)=\frac{1}{8\pi^2}\int_0^\infty\left(\sum_{j=0}^2c_je^{-sm_j^2}\right)\left(sa\coth\left(sa\right)-1\right)\frac{\diff s}{s^3}
    \]
    and
    \begin{equation}\label{eq:T-PV-magnetic-EH}
        f^T_\mathrm{PV}\left(a,\beta\right)=\frac{1}{4\pi^2}\int_0^\infty\sum_{j=0}^2 c_j e^{-sm_j^2}\left(sa\coth\left(sa\right)-1\right)\sum_{n=1}^\infty \left(-1\right)^n e^{-\frac{\beta^2 n^2}{4s}}\frac{\diff s}{s^3}.
    \end{equation}
\end{thm}
The additional regularity needed for this result is expected to be completely technical, as explained in \cite[Remark~3]{GraLewSer-2018-JMPA}. The function $f_\mathrm{PV}^0$ is the well known Euler-Heisenberg energy function at null temperature, which has been introduced for the first time in \cite{HeiEul-1936-ZFP}. Actually, the authors there derive a concave-decreasing and negative function given by \eqref{eq:physics-magnetic-EH} and diverging faster than $\abs{B}^2$ at infinity. For this reason, the total energy of the system, given in the limit by an effective local Lagrangian, is unbounded from below and the model is unstable. Then the same function has been mathematically studied in the same regime of slowly varying purely magnetic fields in \cite{GraLewSer-2018-JMPA}, where the authors derive a non-negative Euler-Heisenberg energy function which guarantees the corresponding model to be stable. Moreover, they find out that the instability of the previous Lagrangian was due to the charge renormalisation \cite[Section~2.3]{GraLewSer-2018-JMPA}. On the other hand, the function $f_\mathrm{PV}^T$ is the non-zero temperature contribution to the total Euler-Heisenberg energy function. To our knowledge, this is the first rigorous derivation of this term, which can be found in the physics literature in \cite{Dit-1979-PRD} and \cite[Equation~(3.162)]{DitGie-2000-book}.
\begin{rmk}\label{rmk:zero-temperature-renormalisation-formula}
    Notice that the thermal contributions in \cref{stat:main-thm-1} and \cref{stat:main-thm-2} (respectively, $M^T$ and $f^T_\mathrm{PV}$) are convergent in the ultraviolet limit (that is, when $m_1,m_2$ go to $+\infty$), meaning that none of these terms is divergent in the high-frequencies domain. This suggests that the well known UV divergences are completely independent on the temperature and arise from the zero temperature theory. Nevertheless, the virtual particles which are used to regularise the free energy functional are assumed to be at the same temperature as the system. So even if the UV divergences are independent on the temperature, the employed subtractions depends on the temperature. This seems to introduce an extra dependance on the temperature which should not be present in the physical expectation values of the free energy\footnote{I would like to thank the anonymous referee for the interesting comment which helped to improve the manuscript.}. However, the regularisation at positive temperature can be rigorously proved to coincide with the zero-temperature regularisation in the large-mass limit. We refer the interested reader to \cref{sec:appendix-A} for all the mathematical details.\\
    In conclusion, the final results of this paper could be made independent on the regularisation parameters by means of the same renormalisation procedure studied at null temperature \cite[Section~2.3]{GraLewSer-2018-JMPA}.
\end{rmk}
Notice also that \cref{eq:T-PV-magnetic-EH} is slightly different from \cref{eq:T-physics-magnetic-EH} appearing in \cite{Dit-1979-PRD}. Indeed, the thermal Euler-Heisenberg formula found in the physics literature is not a well defined object because of the well known ultraviolet divergences in QED. As previously explained, here we deal with this issue by using the Pauli-Villars regularisation technique. In our framework, it correponds to the appearance of the sum in $j$ and consequently of the masses $m_j$'s. Moreover, the cotangent in \eqref{eq:T-physics-magnetic-EH} makes the integral obviously ill-defined due to its infinitely many poles. Thus, we reckon that Dittrich's result should be interpreted as follows: after taking an integration path which contours the poles in order to avoid them, the presence of the imaginary exponential in \eqref{eq:T-physics-magnetic-EH} suggests to rotate this integration domain, leading to the replacement of the cotangent by the hyperbolic cotangent and getting a formula closer to ours in the UV limit. This procedure is standard in the physics literature \cite{Sch-1951-PR, Dun-2005-book}, but of course should be carefully justified from a mathematical point of view. The proof of \cref{stat:main-thm-2} can be found in \cref{sec:the-regime-of-slowly-varying-magnetic-fields}.

\section{Rigorous definition of the PV-regularised free energy functional}\label{sec:rigorous-definition-of-the-PV-regularised-free-energy-functional}
\subsection{Integrable magnetic potentials: proof of the first part of \cref{stat:main-thm-1}}\label{subsec:integrable-magnetic-potentials}
Our starting point is the following formula:
\begin{align}\label{eq:tanh-square}
    x\tanh{x}&=\sum_{k\in\Z}\frac{4x^2}{\left(2k-1\right)^2\pi^2+4x^2}\\ \label{eq:tanh-linear}
    &=\frac{1}{2}\sum_{k\in\Z}\left(2-\frac{i\left(2k-1\right)\pi}{2x+i\left(2k-1\right)\pi}+\frac{i\left(2k-1\right)\pi}{2x-i\left(2k-1\right)\pi}\right).
\end{align}
When $S$ is a self-adjoint operator on $L^2\left(\R^3,\R^4\right)$, with domain $D\left(S\right)$, it follows from \eqref{eq:tanh-linear} using standard functional calculus \cite{ReeSim-1972-book} that the hyberbolic tangent $\tanh{S}$ of $S$ satisfies
\[
S\tanh{S}=\frac{1}{2}\sum_{k\in\Z}\left(2-\frac{i\left(2k-1\right)\pi}{2S+i\left(2k-1\right)\pi}+\frac{i\left(2k-1\right)\pi}{2S-i\left(2k-1\right)\pi}\right).
\]
Notice that this series is convergent when seen as an operator from $D\left(S^2\right)$ to the ambient Hilbert space. Indeed,
\[
\norm{\frac{4S^2}{4S^2+\left(2k-1\right)^2\pi^2}}_{D\left(S^2\right)\xrightarrow{}L^2\left(\R^3,\R^4\right)}\leq\min\left\{1,\left(\left(2k-1\right)\pi\right)^{-2}\norm{4S^2}_{D\left(S^2\right)\xrightarrow{}L^2\left(\R^3,\R^4\right)}\right\}.
\]
Since the domain of $\left(D^0_{m_j}\right)^2$ and $\left(D^A_{m_j}\right)^2$ are both equal to $H^2\left(\R^3,\C^4\right)$, we can write
\begin{multline}\label{eq:T-A}
    T_A\left(\beta\right)=\frac{1}{2\beta}\sum_{k\in\Z}\sum_{j=0}^2 c_j\left(\frac{i\omega\left(k,\beta\right)}{D^A_{m_j}+i\omega\left(k,\beta\right)}-\frac{i\omega\left(k,\beta\right)}{D^A_{m_j}-i\omega\left(k,\beta\right)}\right.\\
    \left.-\frac{i\omega\left(k,\beta\right)}{D^0_{m_j}+i\omega\left(k,\beta\right)}+\frac{i\omega\left(k,\beta\right)}{D^0_{m_j}-i\omega\left(k,\beta\right)}\right),
\end{multline}
on $H^2\left(\R^3,\C^4\right)$, where
\[
\omega\left(k,\beta\right)=\frac{\left(2k-1\right)\pi}{\beta}.
\]
\cref{stat:Dirac-spectral-properties} guarantees us that $0$ is not an eigenvalue of the magnetic Dirac operator $D^A_m$. In any case, this would not have been a problem since $\omega\left(k,\beta\right)\neq 0$, $\forall k\in\Z$.\\
In order to prove the theorem, we want to prove that the the series in the right-hand side of \eqref{eq:T-A} defines a trace class operator whose Schatten norm can be estimated as follows:
\begin{multline*}
    \sum_{k\in\Z}\left\lVert\sum_{j=0}^2 c_j \left(\frac{i\omega\left(k,\beta\right)}{D^A_{m_j}+i\omega\left(k,\beta\right)}-\frac{i\omega\left(k,\beta\right)}{D^A_{m_j}-i\omega\left(k,\beta\right)}\right.\right.\\
    \left.\left.-\frac{i\omega\left(k,\beta\right)}{D^0_{m_j}+i\omega\left(k,\beta\right)}+\frac{i\omega\left(k,\beta\right)}{D^0_{m_j}-i\omega\left(k,\beta\right)}\right)\right\rVert_{\schatten{1}}=\bigO{\beta},
\end{multline*}
as $\beta\xrightarrow{}0$, which can be proved when $A\in L^1\left(\R^3,\R^3\right)\cap H^1\left(\R^3,\R^3\right)$. Indeed, it is not important that $\mathrm{div}\,A=0$ for this result and consequently we do not require $A$ to be in $\Cgaugesobolev$.\\
In order to do so, we just need to derive estimates in Schatten spaces of the operator
\begin{multline*}
    \mathcal{R}\left(\omega\left(k,\beta\right),A\right)=\sum_{j=0}^2 c_j \left(\frac{i\omega\left(k,\beta\right)}{D^A_{m_j}+i\omega\left(k,\beta\right)}-\frac{i\omega\left(k,\beta\right)}{D^A_{m_j}-i\omega\left(k,\beta\right)}\right.\\
    \left.-\frac{i\omega\left(k,\beta\right)}{D^0_{m_j}+i\omega\left(k,\beta\right)}+\frac{i\omega\left(k,\beta\right)}{D^0_{m_j}-i\omega\left(k,\beta\right)}\right),
\end{multline*}
which we can then sum with respect to $k\in\Z$. To this end, by using the resolvent formula
\[
\frac{i\omega\left(k,\beta\right)}{D^A_{m_j}+i\omega\left(k,\beta\right)}-\frac{i\omega\left(k,\beta\right)}{D^0_{m_j}+i\omega\left(k,\beta\right)}=\frac{i\omega\left(k,\beta\right)}{D^A_{m_j}+i\omega\left(k,\beta\right)}\left(\boldsymbol{\alpha}\cdot A\right)\frac{1}{D^0_{m_j}+i\omega\left(k,\beta\right)}
\]
and iterating it six times, we get
\begin{multline*}
    \mathcal{R}\left(\omega\left(k,\beta\right),A\right)=\sum_{n=1}^5 \left(R_n\left(\omega\left(k,\beta\right),A\right)+R_n\left(-\omega\left(k,\beta\right),A\right)\right)\\
    +\left(R'_6\left(\omega\left(k,\beta\right),A\right)+R'_6\left(-\omega\left(k,\beta\right),A\right)\right),
\end{multline*}
with
\begin{equation}\label{eq:R-n}
    R_n\left(\omega\left(k,\beta\right),A\right)=\sum_{j=0}^2 c_j\frac{i\omega\left(k,\beta\right)}{D^0_{m_j}+i\omega\left(k,\beta\right)}\left(\left(\boldsymbol{\alpha}\cdot A\right)\frac{1}{D^0_{m_j}+i\omega\left(k,\beta\right)}\right)^n
\end{equation}
and
\begin{equation}\label{eq:R'-6}
    R'_6\left(\omega\left(k,\beta\right),A\right)=\sum_{j=0}^2 c_j\frac{i\omega\left(k,\beta\right)}{D^A_{m_j}+i\omega\left(k,\beta\right)}\left(\left(\boldsymbol{\alpha}\cdot A\right)\frac{1}{D^0_{m_j}+i\omega\left(k,\beta\right)}\right)^6.
\end{equation}
Basically, by replacing the integral with respect to $\omega$ by the series with respect to $k$, the same proof as in \cite[Theorem~2.1]{GraHaiLewSer-2013-ARMA} holds giving that
\begin{multline*}
    \sum_{k\in\Z}\left(\sum_{j=1}^5\norm{R_n\left(\omega\left(k,\beta\right),A\right)+R_n\left(-\omega\left(k,\beta\right),A\right)}_{\schatten{1}}\right.\\
    \left.+\norm{R'_6\left(\omega\left(k,\beta\right),A\right)+R'_6\left(-\omega\left(k,\beta\right),A\right)}_{\schatten{1}}\right)=\bigO{\beta},
\end{multline*}
as $\beta\xrightarrow{}0$. In particular, the dependence of $\bigO{\beta}$ on the magnetic potential $A$ occurs through the $L^p$-norms of it, for $p\in\intinteger{1}{5}$, as well as the $L^2$-norm of its associated magnetic field. Therefore, these estimates are not gauge-invariant and this fact justifies the following subsection.\qed
\begin{rmk}
    Notice that at this stage it could have been possible to include an external electric potential as well. Indeed, the same estimates as the ones in \cite[Proof~of~Proposition~3.1]{GraHaiLewSer-2013-ARMA} hold as long as the $\C^4$-trace is applied before taking the trace over $L^2\left(\R^3,\C\right)$ (see the comment just after \cite[Theorem~2.1]{GraHaiLewSer-2013-ARMA} for further details). Nevertheless, we are not able to prove the results in the following subsection in the complete electromagnetic case, mainly because of the lack of the integration by parts due to the discretisation of $\omega$ (integrals in $d\omega$ are replaced by series over a discrete set of $\omega$ at finite temperature). Moreover, a rigorous derivation of the Euler-Heisenberg formula in presence of an electric field is a really complicated task for the moment and exists not even at null temperature. For all of these reasons, we decided to restrict this paper to the purely magnetic case since the beginning, as mentioned in the introduction.
\end{rmk}

\subsection{Gauge-invariant estimates}\label{subsec:gauge-invariant-estimates}
In \cref{subsec:integrable-magnetic-potentials}, we showed that the operator
\[
T_A\left(\beta\right)=\frac{1}{2}\sum_{j=0}^2 c_j\tr_{\C^4}\left(D^0_{m_j}\tanh{\frac{\beta D^0_{m_j}}{2}}-D^A_{m_j}\tanh{\frac{\beta D^A_{m_j}}{2}}\right)
\]
is trace class on $L^2\left(\R^3,\C^4\right)$ when the magnetic potential $A$ decays fast enough, namely when $A\in L^1\left(\R^3,\R^3\right)\cap H^1\left(\R^3,\R^3\right)$. More precisely, in the proof of the first part of \cref{stat:main-thm-1}, we write
\begin{align*}
    T_A\left(\beta\right)=\sum_{n=1}^5 T_n\left(A,\beta\right)+T'_6\left(A,\beta\right)=&\frac{1}{2\beta}\sum_{n=1}^5\sum_{k\in\Z}\left(R_n\left(\omega\left(k,\beta\right),A\right)+R_n\left(-\omega\left(k,\beta\right),A\right)\right)\\
    &+\frac{1}{2\beta}\sum_{k\in\Z}\left(R'_6\left(\omega\left(k,\beta\right),A\right)+R'_6\left(-\omega\left(k,\beta\right),A\right)\right),
\end{align*}
with $R_n$, $R'_6$ given by \eqref{eq:R-n} and \eqref{eq:R'-6}, and we prove that the operators $T_n\left(A,\beta\right)$ and $T'_6\left(A,\beta\right)$ are trace class. Nevertheless, our estimates involve non gauge-invariant quantities (some $L^p$ norms of $A$) and they require that $A$ decays fast enough at infinity. We are interested in estimates which involve only the norm $\norm{\mathrm{curl}A}_{L^2}$. Notice that in the previous section it was not essential that $\mathrm{div}A=0$. We will need this hypothesis later in this work (see \cref{stat:explicit-second-order} below).\\
First of all, let us state the following lemma which says that the trace of the odd order operators $T_1\left(A,\beta\right)$, $T_3\left(A,\beta\right)$ and $T_5\left(A,\beta\right)$ vanishes. This is a consequence of the charge-conjugation invariance which is usually known as Furry's theorem in the physics literature (see \cite{Fur-1937-PR} and \cite[Sec.~4.1]{GreRei-2009-book}).
\begin{lem}\label{stat:furry-odd-order}
    For $A\in L^1\left(\R^3,\R^3\right)\cap H^1\left(\R^3,\R^3\right)$ and $n=1,3,5$, we have
    \begin{multline*}
        \frac{1}{\beta}\int_0^\beta\tr\left[T_n\left(A,b\right)\right]\diff b\\
        =\frac{1}{2\beta}\int_0^\beta\sum_{k\in\Z}\tr\left[R_n\left(\omega\left(k,b\right),A\right)+R_n\left(-\omega\left(k,b\right),A\right)\right]\frac{\diff b}{b}=0.
    \end{multline*}
\end{lem}
\proof
The proof is the same as in \cite[Lemma~4.1]{GraHaiLewSer-2013-ARMA}.
\endproof
We now compute exactly the second order term $\mathcal{F}_{\mathrm{PV},2}$ associated to the term $T_2\left(A,\beta\right)$ appearing in the decomposition of $T_A\left(\beta\right)$, assuming that $A\in H^1\left(\R^3,\R^3\right)$ and $\mathrm{div}A=0$.
\begin{lem}\label{stat:explicit-second-order}
    For $A\in L^2\left(\R^3,\R^3\right)\cap\Cgaugesobolev$, we have
    \begin{equation}\label{eq:vacuum-response}
        \frac{1}{\beta}\int_0^\beta\tr\left[T_2\left(A,b\right)\right]\diff b=\frac{1}{8\pi}\int_{\R^3}\left(M^0\left(q\right)+M^T\left(q,\beta\right)\right)\abs{\widehat{B}\left(q\right)}^2\diff q=\mathcal{F}_{\mathrm{PV},2}\left(B,\beta\right),
    \end{equation}
    with
    \begin{equation}\label{eq:0-vacuum-response}
        M^0\left(q\right)=-\frac{2}{\pi}\int_0^1\sum_{j=0}^2 c_j\log\left(m_j^2+u\left(1-u\right)q^2\right)u\left(1-u\right)\diff u
    \end{equation}
    and
    \begin{multline}\label{eq:T-vacuum-response}
        M^T\left(q,\beta\right)=-\frac{8}{\pi}\int_0^1\int_0^\infty\sum_{j=0}^2 c_j\left[\frac{1}{1+e^{-X_j\left(\beta,u,q\right)\cosh t}}+\frac{1}{X_j\left(\beta,u,q\right)\cosh t}\left(2\log 2\right.\right.\\
        \left.\left.-\log\left(\left(e^{X_j\left(\beta,u,q\right)\cosh t}+1\right)\left(e^{-X_j\left(\beta,u,q\right)\cosh t}+1\right)\right)\right)\right]\diff t\,u\left(1-u\right)\diff u,
    \end{multline}
    where $X_j\left(b,u,q\right)=b\sqrt{m_j^2+u\left(1-u\right)q^2}$ for $j=0,1,2$.
\end{lem}
The function $M^0$ in formula \eqref{eq:vacuum-response} is the dielectric response of Dirac's vacuum at zero temperature which has already been rigorously computed in \cite[Formula~(2.21)]{GraHaiLewSer-2013-ARMA} and is well known in the physics literature \cite[Eq.~(5.39)]{GreRei-2009-book}. On the other hand, $M^T$ defined in \eqref{eq:T-vacuum-response} is the contribution at finite temperature $T$ to the dielectric response of Dirac's vacuum. Notice that the integral with respect to $t$ in \eqref{eq:T-vacuum-response} is perfectly defined and converges for $q=0$, namely $M^T$ is bounded in a neighborhood of $q=0$. This means that we cannot expect an effect similar to the one shown in \cite{HaiLewSei-2008-RMP} for the reduced BDF model, where the authors prove that due to the positive temperature the particles in the polarised Dirac sea rearrange themselves so as to completely screen the external potential. As already mentioned, this phenomenon is known as Debye screening in non-relativistic fermionic plasma physics. After all, this could be explained by the nature of the magnetic field which basically acts on charged particles by confining them in a space region rather than attracting or repelling them.
\proof
First of all, by comparing the calculations in \cref{subsec:integrable-magnetic-potentials} to the calculations at zero temperature \cite{GraHaiLewSer-2012-JEDP}, notice that we simply have to make the following formal replacements in order to include a finite temperature:
\begin{gather*}
    \frac{1}{4\pi}\int_\R\diff\omega\xrightarrow{}\frac{1}{2\beta}\sum_{k\in\Z},\\
    \omega\in\R\xrightarrow{}\omega\left(k,\beta\right)=\frac{\left(2k-1\right)\pi}{\beta},\,k\in\Z,
\end{gather*}
where $\beta=1/T\in\intoo{0}{\infty}$, and then to consider the averaged primitive function with respect to $\beta$ of each term in the expansion. By keeping this prescription in mind, the same computations as in the proof of \cite[Lemma~4.2]{GraHaiLewSer-2013-ARMA} yields
\[
\tr\left[T_2\left(A,\beta\right)\right]=\int_{\R^3}G\left(q,\beta\right)\abs{\widehat{A}\left(q\right)}^2\diff q,
\]
with
\[
G\left(q,\beta\right)=\frac{1}{\beta\pi^3}\int_{\R^3}\sum_{k\in\Z}\sum_{j=0}^2 c_j\frac{\left(p\cdot q\right)\omega\left(k,\beta\right)^2}{\left(p^2+m_j^2+\omega\left(k,\beta\right)^2\right)^2\left(\left(p-q\right)^2+m_j^2+\omega\left(k,\beta\right)^2\right)}\diff p.
\]
We then use the identity (see \cite[Chap.~5]{GreRei-2009-book})
\[
\frac{1}{a^2 b}=\int_0^1\left(\int_0^\infty s^2 e^{-s\left(ua+\left(1-u\right)b\right)}\diff s\right)u\diff u
\]
to rewrite
\begin{multline*}
    G\left(q,\beta\right)=\frac{1}{\beta\pi^3}\int_0^1\int_0^\infty\sum_{j=0}^2 c_j e^{-s\left(m_j^2+\left(1-u\right)q^2\right)} \left(\sum_{k\in\Z}e^{-s\omega\left(k,\beta\right)^2}\omega\left(k,\beta\right)^2\right) \\
    \times \int_{\R^3}\left(p\cdot q\right)e^{-s\left(p^2-2\left(1-u\right)p\cdot q\right)}\diff p\,s^2 \diff s\,u\diff u.
\end{multline*}
This is justified by conditions \eqref{eq:PV-conditions} which allow us to apply Fubini's theorem to recombine the integrals. Since
\[
\sum_{k\in\Z}e^{-s\omega\left(k,\beta\right)^2}\omega\left(k,\beta\right)^2=\sum_{k\in\Z}e^{-s\frac{4\pi^2}{\beta^2}\left(k-\frac{1}{2}\right)^2}\frac{4\pi^2}{\beta^2}\left(k-\frac{1}{2}\right)^2=-\frac{\diff}{\diff s}\theta_2\left(0,\frac{4\pi is}{\beta^2}\right),
\]
where $\theta_2$ is a Jacobi theta function defined by
\begin{equation}\label{eq:jacobi-theta-2}
    \theta_2\left(z,\tau\right)=e^{i\tau \frac{\pi}{4}+i\pi z}\theta\left(z+\frac{\tau}{2},\tau\right)=\sum_{n=-\infty}^{+\infty}e^{i\pi z\left(2n+1\right)}e^{i\pi\tau\left(n+\frac{1}{2}\right)^2},
\end{equation}
with
\[
\theta\left(z,\tau\right)=\sum_{n=-\infty}^{+\infty}e^{2\pi i nz+\pi i n^2 \tau}
\]
(to our knowledge, it does not exist an unambiguous definition in the literature, see \cite[Chap.~16]{AbrSte-1964-book} and \cite{Bel-1961-book} for example), and
\begin{align*}
    \int_{\R^3}\left(p\cdot q\right)e^{-s\left(p^2-2\left(1-u\right)p\cdot q\right)}\diff p&=q\cdot\nabla_x\left(\int_{\R^3}e^{p\cdot x-sp^2+2s\left(1-u\right)p\cdot q}\diff p\right)\bigg|_{x=0}\\
    &=\left(\frac{\pi}{s}\right)^{3/2}\left(1-u\right)q^2e^{s\left(1-u\right)^2q^2},
\end{align*}
we get
\begin{equation}\label{eq:G-function}
    G\left(q,\beta\right)=-\frac{q^2}{\beta\pi^{3/2}}\int_0^1\int_0^\infty\sum_{j=0}^2 c_j\left(\frac{\diff}{\diff s}\theta_2\left(0,\frac{4\pi is}{\beta^2}\right)\right)s^{1/2}e^{-s\left(m_j^2+u\left(1-u\right)q^2\right)}\diff s\,u\left(1-u\right)\diff u.
\end{equation}
Integrating by parts, we then get
\begin{multline*}
    G\left(q,\beta\right)=\frac{q^2}{\beta\pi^{3/2}}\int_0^1\int_0^\infty \sum_{j=0}^2 c_j\theta_2\left(0,\frac{4\pi is}{\beta^2}\right)e^{-s\left(m_j^2+u\left(1-u\right)q^2\right)}\\
    \times\left(\frac{1}{2s^{1/2}}-s^{1/2}\left(m_j^2+u\left(1-u\right)q^2\right)\right)\diff s\,u\left(1-u\right)\diff u.
\end{multline*}
We now want to properly normalise the function $G\left(q\right)$ so as to reproduce the dielectric response of Dirac's vacuum at null temperature when $T$ goes to zero. As shown by Dittrich in \cite{Dit-1979-PRD}, this can be done by extracting the $n=0$ term in the series defining the function $\theta_2$ above:
\begin{equation}\label{eq:theta-function-poisson}
    \theta_2\left(0,\frac{4\pi is}{\beta^2}\right)=\sum_{n\in\Z}e^{-s\frac{4\pi^2}{\beta^2}\left(n-\frac{1}{2}\right)^2}=\left(\frac{1}{\pi s}\right)^{1/2}\frac{\beta}{2}\left[1+\sum_{n\neq 0}e^{-\frac{\beta^2}{4s}n^2}e^{-i\pi n}\right],    
\end{equation}
where in the second equality we first use Poisson summation formula \cite[Formula~6.5]{Bel-1961-book} and then extract the $n=0$ term. Thus, our result can be expressed so far as
\[
G\left(q,\beta\right)=G^0\left(q\right)+G^T\left(q,\beta\right),
\]
where
\[
G^0\left(q\right)=\frac{q^2}{2\pi^2}\int_0^1\int_0^\infty\sum_{j=0}^2 c_j e^{-s\left(m_j^2+u\left(1-u\right)q^2\right)}\left(\frac{1}{2s}-\left(m_j^2+u\left(1-u\right)q^2\right)\right)\diff s\,u\left(1-u\right)\diff u
\]
is the null temperature term, and
\begin{multline*}
    G^T\left(q,\beta\right)=\frac{q^2}{\beta\pi^{3/2}}\int_0^1\int_0^\infty \sum_{j=0}^2 c_j\left(\theta_2\left(0,\frac{4\pi is}{\beta^2}\right)-\left(\frac{1}{\pi s}\right)^{1/2}\frac{\beta}{2}\right)e^{-s\left(m_j^2+u\left(1-u\right)q^2\right)}\\
    \times\left(\frac{1}{2s^{1/2}}-s^{1/2}\left(m_j^2+u\left(1-u\right)q^2\right)\right)\diff s\,u\left(1-u\right)\diff u
\end{multline*}
contains the non-zero temperature contribution. Let us now focus on the function $G^0$. Integrating by parts and again thanks to conditions \eqref{eq:PV-conditions}, we have
\begin{multline*}
    \int_0^\infty\sum_{j=0}^2 c_j e^{-s\left(m_j^2+u\left(1-u\right)q^2\right)}s^{-1}\diff s\\
    =\int_0^\infty\sum_{j=0}^2 c_j \log\left(s\right) e^{-s\left(m_j^2+u\left(1-u\right)q^2\right)}\left(m_j^2+u\left(1-u\right)q^2\right)\diff s.
\end{multline*}
The change of variables $\sigma=s\left(m_j^2+u\left(1-u\right)q^2\right)$ yields
\begin{align*}
    \int_0^\infty\sum_{j=0}^2 c_j e^{-s\left(m_j^2+u\left(1-u\right)q^2\right)}s^{-1}\diff s&=\int_0^\infty \sum_{j=0}^2 c_j \log\left(\frac{\sigma}{m_j^2+u\left(1-u\right)q^2}\right)e^{-\sigma}\diff\sigma\\
    &=-\sum_{j=0}^2 c_j \log\left(m_j^2+u\left(1-u\right)q^2\right),
\end{align*}
and
\[
\int_0^1\int_0^\infty\sum_{j=0}^2 c_j e^{-\sigma}\diff\sigma\,u\left(1-u\right)\diff u=-\int_0^1\sum_{j=0}^2 c_j u\left(1-u\right)\diff u=0,
\]
thanks to conditions \eqref{eq:PV-conditions}. Finally, we get
\[
G^0\left(q\right)=-\frac{q^2}{4\pi^2}\int_0^1\sum_{j=0}^2 c_j \log\left(m_j^2+u\left(1-u\right)q^2\right)u\left(1-u\right)\diff u=\frac{q^2}{8\pi}M^0\left(q\right).
\]
Let us now focus on $G^T$ which can be written as follows:
\begin{multline*}
    G^T\left(q,\beta\right)=\frac{q^2}{2\pi^2}\sum_{n\neq 0}\left(-1\right)^n\int_0^1\int_0^\infty \sum_{j=0}^2 c_j e^{-s\left(m_j^2+u\left(1-u\right)q^2\right)-\frac{\beta^2 n^2}{4}\frac{1}{s}}\\
    \times\left(\frac{1}{2s}-\left(m_j^2+u\left(1-u\right)q^2\right)\right)\diff s\,u\left(1-u\right)\diff u
\end{multline*}
We then use the identity (see \cite[Formulas~(3.324(1))~and~(3.471(9))]{GraRyz-1965-book})
\[
\int_{0}^\infty x^{\nu-1}\exp{\left(-\alpha\frac{1}{x}-\gamma x\right)}\diff x=2\left(\frac{\alpha}{\gamma}\right)^{\nu/2}K_\nu\left(2\sqrt{\alpha\nu}\right),\quad\mathrm{Re}\beta,\mathrm{Re}\gamma>0,
\]
to compute
\[
\int_0^\infty\sum_{j=0}^2 c_j \frac{1}{2s}e^{-s\left(m_j^2+u\left(1-u\right)q^2\right)-\frac{\beta^2 n^2}{4}\frac{1}{s}}\diff s=\sum_{j=0}^2 c_j K_0\left(\beta n\sqrt{m_j^2+u\left(1-u\right)q^2}\right)
\]
and
\begin{multline*}
    \int_0^\infty\sum_{j=0}^2 c_j e^{-s\left(m_j^2+u\left(1-u\right)q^2\right)-\frac{\beta^2 n^2}{4}\frac{1}{s}}\left(m_j^2+u\left(1-u\right)q^2\right)\diff s\\
    =\sum_{j=0}^2c_j \beta n\sqrt{m_j^2+u\left(1-u\right)q^2}K_1\left(\beta n\sqrt{m_j^2+u\left(1-u\right)q^2}\right),
\end{multline*}
where $K_0$ and $K_1$ are modified Bessel functions of second kind \cite[Chap.~9]{AbrSte-1964-book}. Finally, by using the parity of Bessel functions, we get:
\begin{multline*}
    G^T\left(q,\beta\right)=\frac{q^2}{\pi^2}\sum_{n=1}^\infty\left(-1\right)^n\int_0^1\sum_{j=0}^2 c_j\left[K_0\left(\beta n\sqrt{m_j^2+u\left(1-u\right)q^2}\right)\right.\\
    \left.-\beta n\sqrt{m_j^2+u\left(1-u\right)q^2}K_1\left(\beta n\sqrt{m_j^2+u\left(1-u\right)q^2}\right)\right]u\left(1-u\right)\diff u.
\end{multline*}
Now, by exploiting the Sommerfeld integral representation \cite[pp.~328]{NikOuv-1983-book} for Bessel functions of second kind,
\[
K_\nu\left(x\right)=\int_0^\infty e^{-x\cosh t}\cosh\left(\nu t\right)\diff t,\;x>0,
\]
and by rearranging the integrals and the series thanks to conditions \eqref{eq:PV-conditions}, we get
\begin{multline*}
    G^T\left(q,\beta\right)=\frac{q^2}{\pi^2}\int_0^1\int_0^\infty\sum_{j=0}^2\sum_{n=1}^\infty\left(-1\right)^n c_j \left(e^{-n X_j\left(\beta,u,q\right)\cosh t}\right.\\
    \left.-n X_j\left(\beta,u,q\right)\cosh t\,e^{-n X_j\left(\beta,u,q\right)\cosh t}\right)\diff t\,u\left(1-u\right)\diff u,   
\end{multline*}
where $X_j\left(\beta,u,q\right)=\beta\sqrt{m_j^2+u\left(1-u\right)q^2}$ for $j=0,1,2$. Finally, the identities
\[
\sum_{n=1}^\infty\left(-e^{-X_j\left(\beta,u,q\right)\cosh t}\right)^n= -\frac{e^{-X_j\left(\beta,u,q\right)\cosh t}}{1+e^{-X_j\left(\beta,u,q\right)\cosh t}}
\]
and
\begin{multline*}
    \sum_{n=1}^\infty n X_j\left(\beta,u,q\right)\cosh t\left(-e^{-X_j\left(\beta,u,q\right)\cosh t}\right)^n\\
    =X_j\left(\beta,u,q\right)\cosh t\,e^{-X_j\left(\beta,u,q\right)\cosh t}\frac{\diff}{\diff y}\frac{y}{1+y}\bigg|_{y=e^{-X_j\left(\beta,u,q\right)\cosh t}}\\
    =X_j\left(\beta,u,q\right)\cosh t\frac{e^{-X_j\left(\beta,u,q\right)\cosh t}}{\left(1+e^{-X_j\left(\beta,u,q\right)\cosh t}\right)^2}
\end{multline*}
yield
\begin{multline}\label{eq:original-formula-G-T}
    G^T\left(q,\beta\right)=-\frac{q^2}{\pi^2}\int_0^1\int_0^\infty\sum_{j=0}^2 c_j\frac{e^{-X_j\left(\beta,u,q\right)\cosh t}\left(1+X_j\left(\beta,u,q\right)\cosh t+e^{-X_j\left(\beta,u,q\right)\cosh t}\right)}{\left(1+e^{-X_j\left(\beta,u,q\right)\cosh t}\right)^2}\\
    \times dt\,u\left(1-u\right)\diff u.
\end{multline}
Now,
\begin{multline*}
    \frac{1}{\beta}\int_0^\beta G^T\left(q,b\right)\diff b=-\frac{q^2}{\pi^2}\int_0^1\int_0^\infty\frac{1}{\beta}\int_0^\beta\sum_{j=0}^2 c_j e^{-X_j\left(\beta,u,q\right)\cosh t}\\
    \times \frac{\left(1+X_j\left(\beta,u,q\right)\cosh t+e^{-X_j\left(\beta,u,q\right)\cosh t}\right)}{\left(1+e^{-X_j\left(\beta,u,q\right)\cosh t}\right)^2}\diff t\,u\left(1-u\right)\diff u
\end{multline*}
and integrating with respect to $\beta$ we get
\begin{multline*}
    \frac{1}{\beta}\int_0^\beta G^T\left(q,b\right)\diff b=-\frac{q^2}{\pi^2}\int_0^1\int_0^\infty\sum_{j=0}^2 c_j\left[\frac{1}{1+e^{-X_j\left(\beta,u,q\right)\cosh t}}+\frac{1}{X_j\left(\beta,u,q\right)\cosh t}\left(2\log 2\right.\right.\\
    \left.\left.-\log\left(\left(e^{X_j\left(\beta,u,q\right)\cosh t}+1\right)\left(e^{-X_j\left(\beta,u,q\right)\cosh t}+1\right)\right)\right)\right]\diff t\,u\left(1-u\right)\diff u.
\end{multline*}
Remarking that
\[
\frac{1}{\beta}\int_0^\beta G^T\left(q,b\right)\diff b=\frac{q^2}{8\pi}M^T\left(q,\beta\right),  
\]
allows us to conclude the proof.
\endproof
\begin{rmk}\label{stat:KMS-condition}
    The formal replacement presented at the beginning of the proof is strictly related to the KMS condition which prescribes the anti-periodicity with respect to an imaginary time for Green's functions. However, this substitution is derived here by following a completely different argument, which is purely mathematical. More specifically, the replacement of integrals over frequencies by series basically arises from the representation of $x\tanh x$ as a power series.
\end{rmk}
\begin{rmk}
    Notice that by \eqref{eq:theta-function-poisson} it is easy to see that $G^T$ goes to zero as $T$ goes to zero, or equivalently $\beta$ goes to infinity, so as to recover the well known formula for the dielectric response of Dirac's vacuum at null temperature \cite[Lemma~4.2]{GraHaiLewSer-2013-ARMA}.
\end{rmk}
\begin{rmk}
    In our framework, by doing the same calculations with an electric potential $V$, we get a different function $G$ since the beginning. By looking at the limit of this function as $\beta$ goes to infinity and replacing the Riemann sum by an integral, we recover the same result found in \cite[Formula~(4.16)]{GraHaiLewSer-2013-ARMA} in the null temperature case.
\end{rmk}
\begin{rmk}
    In order to convince the reader that our framework is coherent with what the authors have done in \cite{HaiLewSei-2008-RMP}, let us come back to the functional $\mathcal{F}$ defined in \eqref{eq:infinite-free-energy}. In presence of a purely electrostatic potential, instead of applying the Pauli-Villars regularisation technique, we could introduce a sharp cut-off by replacing the trace over the whole space by a trace in the space of operators acting on functions whose Fourier transform is compactly supported in a ball of radius $\Lambda$,
    \[
    \mathcal{F}_\Lambda\left(V,\beta\right)=-\frac{1}{\beta}\tr_\Lambda\left[\log\left(2\cosh\frac{\beta D^V_m}{2}\right)\right].
    \]
    By adding the infinite constant $\left(1/\beta\right)\tr_\Lambda\left[\log\left(2\cosh \beta D^0_m/2\right)\right]$, we get a cut-off-regularised free energy functional. The $\Lambda-$trace results in integrals over $B\left(0,\Lambda\right)$ in the momentum space and our calculations reduce to the ones in \cite{HaiLewSei-2008-RMP}. Thus, at non-zero temperature we get a response of Dirac's vacuum to the magnetic field different from the one to the electric potential, whereas it is known to be the same at null temperature. This explains once more why an effect similar to the Debye screening appears only in presence of an electric field.
\end{rmk}
In order to complete the study of the second order term, the following lemma states the main properties of the non-zero temperature contribution $M^T$ to the response function:
\begin{lem}\label{stat:MT-properties}
    The function $M^T$ given by \eqref{eq:T-vacuum-response} is well defined, bounded for any fixed $\beta\in\intoo{0}{+\infty}$ and such that
    \begin{equation*}
        M^0\left(q\right)+M^T\left(q,\beta\right)\geq 0,
    \end{equation*}
    where $M^0$ is given by \eqref{eq:0-vacuum-response}.
\end{lem}
\proof
We start remarking that $M^T$ can be rewritten by means of \cref{eq:original-formula-G-T} as:
\begin{multline*}
    M^T\left(q,\beta\right)=-\frac{8}{\pi\beta}\int_0^1\int_0^\infty\int_0^\beta\sum_{j=0}^2 c_j\frac{e^{-X_j\left(b,u,q\right)\cosh t}\left(1+X_j\left(b,u,q\right)\cosh t+e^{-X_j\left(b,u,q\right)\cosh t}\right)}{\left(1+e^{-X_j\left(b,u,q\right)\cosh t}\right)^2}\\
    \times \diff b\diff t\,u\left(1-u\right)\diff u,
\end{multline*}
where $X_j\left(b,u,q\right)=b\sqrt{m_j^2+u\left(1-u\right)q^2}$ for $j=0,1,2$. Every term in the Pauli-Villars sum is obviously positive and can be estimated as follows:
\begin{multline*}
    \frac{e^{-X_j\left(b,u,q\right)\cosh t}\left(1+X_j\left(b,u,q\right)\cosh t+e^{-X_j\left(b,u,q\right)\cosh t}\right)}{\left(1+e^{-X_j\left(b,u,q\right)\cosh t}\right)^2}\\
    \leq e^{-X_j\left(b,u,q\right)\cosh t}\left(1+X_j\left(b,u,q\right)\cosh t\right)\\
    \leq e^{-X_j\left(b,u,q\right)\cosh t}+e^{-\frac{1}{2}X_j\left(b,u,q\right)\cosh t}\\
    \leq 2 e^{-\frac{1}{2}X_j\left(b,u,q\right)\cosh t}\\
    \leq 2 e^{-\frac{1}{2}X_j\left(b,u,q\right)\left(1+\frac{t^2}{2}\right)},
\end{multline*}
so that
\begin{multline*}
    \frac{2}{\beta}\int_0^1 u\left(1-u\right)\int_0^\beta e^{-\frac{1}{2}X_j\left(b,u,q\right)}\int_0^\infty e^{-\frac{t^2}{4}X_j\left(b,u,q\right)}\diff t\diff b\diff u\\
    =\frac{\sqrt{\pi}}{3\sqrt{m_j}\beta}\int_0^\beta \frac{1}{\sqrt{X_j\left(b,u,q\right)}}e^{-\frac{1}{2}X_j\left(b,u,q\right)}\diff b\\
    \leq \frac{2\sqrt{\pi}}{3\sqrt{m_j}}\frac{1}{\sqrt{\beta}}.
\end{multline*}
In particular, the previous calculations show that the function $M^T$ is well defined and bounded for any fixed $\beta\in\intoo{0}{+\infty}$. Now, in order to conclude the proof, let us recall that
\[
\frac{q^2}{8\pi}\left(M^0\left(q\right)+M^T\left(q,\beta\right)\right)=\frac{1}{\beta}\int_0^\beta G\left(q,b\right)\diff b
\]
and $G\left(q,\beta\right)\geq 0$ by \cite[Lemma~11]{GraLewSer-2018-JMPA}. Indeed,
\[
\frac{\diff}{\diff s}\theta_2\left(0,\frac{4\pi is}{\beta^2}\right)=-\frac{4\pi^2}{\beta^2}\sum_{n\in\Z}\left(n+\frac{1}{2}\right)^2 e^{-\frac{4\pi^2}{\beta^2}s\left(n+\frac{1}{2}\right)^2}\leq 0
\]
and in particular the function $G$ is given by \eqref{eq:G-function}.
\endproof
\begin{rmk}
    The calculations in the previous proof show that $M^T$ vanishes as $\beta$ goes to infinity as expected since we isolated the zero temperature term $M^0$. In addition, let us notice that the terms corresponding to $c_1$ and $c_2$ goes to zero when $m_1,m_2\xrightarrow{}\infty$, that is in the large cut-off limit (recall formula \eqref{eq:averaged-UV-cutoff}). Since the $j=0$ term is negative, this implies that $M^T$ stays negative for $\Lambda$ large enough.
\end{rmk}

\subsection{Proof of the second part of \cref{stat:main-thm-1}}\label{subsec:proof-of-the-second-part-of-main-thm-1}
First of all, given any $A\in L^1\left(\R^3,\R^3\right)\cap\Cgaugesobolev$, we proved that the functional $\mathcal{F}_\mathrm{PV}$ is well defined and can be written as
\[
\mathcal{F}_\mathrm{PV}\left(A,\beta\right)=\mathcal{F}_{\mathrm{PV},2}\left(B,\beta\right)+\mathcal{R}\left(A,\beta\right),
\]
where
\[
\mathcal{F}_{\mathrm{PV},2}\left(B,\beta\right)=\frac{1}{\beta}\int_0^\beta\tr\left[T_2\left(A,b\right)\right]\diff b
\]
and
\[
\mathcal{R}\left(A,\beta\right)=\frac{1}{\beta}\int_0^\beta\left(\tr\left[T_4\left(A,b\right)\right]+\tr\left[T_6\left(A,b\right)\right]\right)\diff b.
\]
Almost the same proof as in \cite[Theorem~2.1]{GraHaiLewSer-2013-ARMA} holds, by replacing \cite[Lemma~4.1]{GraHaiLewSer-2013-ARMA} by \cref{stat:furry-odd-order} and \cite[Lemma~4.2]{GraHaiLewSer-2013-ARMA} by \cref{stat:explicit-second-order}. In order to include a finite temperature, as already remarked in \cref{stat:explicit-second-order}, recall that we simply have to make the following formal replacements:
\begin{gather*}
    \frac{1}{4\pi}\int_\R\diff\omega\xrightarrow{}\frac{1}{2\beta}\sum_{k\in\Z},\\
    \omega\in\R\xrightarrow{}\omega\left(k,\beta\right)=\frac{\left(2k-1\right)\pi}{\beta},\,k\in\Z
\end{gather*}
where $\beta=1/T\in\intoo{0}{\infty}$, and then to consider the averaged primitive function with respect to $\beta$ of each term in the expansion. After the substitutions, it is important to notice that the universal constant $K$ in \cite[Lemma~4.4]{GraHaiLewSer-2013-ARMA} and \cite[proof of Lemma~4.5]{GraHaiLewSer-2013-ARMA} translates into a positive temperature constant which is $\bigO{\beta}$ as $\beta\xrightarrow{}0$ and this is crucial in order to get the finite constant $\kappa$ in \eqref{eq:remainder-estimate} (recall that one has to pay attention to the singularity of the factor $1/\beta$ because of the integration around zero). Indeed, after replacing the integrals over frequencies by series, for instance for the fourth order term, the positive temperature constant would depend on
\begin{align*}
    \sum_{j=0}^2 \abs{c_j} \sum_{k\in\Z}\abs{\frac{\omega\left(k,\beta\right)^2}{\left(m_j^2+\omega\left(k,\beta\right)^2\right)^2}}&=\beta^2 \sum_{j=0}^2 \abs{c_j} \sum_{k\in\Z}\abs{\frac{\left(2k-1\right)^2 \pi^2}{\left(\beta^2 m_j^2+\left(2k-1\right)^2\pi^2\right)^2}}\\
    &\leq \frac{\beta^2}{\pi^2} \sum_{j=0}^2 \abs{c_j} \sum_{k\in\Z}\abs{\frac{1}{\left(2k-1\right)^2}}\leq C\beta^2,
\end{align*}
which clearly shows the constant to be $\bigO{\beta}$ as $\beta\to 0$. A similar argument works for the sixth order term. Notice also that $0$ is never an eigenvalue of the magnetic Dirac operator by \cref{stat:Dirac-spectral-properties}. Finally, a proof of the properties of $M^T$ can be found in \cref{stat:MT-properties}.\qed

\section{The regime of slowly varying magnetic fields: proof of \cref{stat:main-thm-2}}\label{sec:the-regime-of-slowly-varying-magnetic-fields}
This section is devoted to the proof of \cref{stat:main-thm-2}. In order to do so, we will exploit the general scheme used in the proof of \cite[Proposition~17]{GraLewSer-2018-JMPA}. Thus, let us recall the general setting: we consider a scaled magnetic field $B_\varepsilon\left(x\right)=B\left(\varepsilon x\right)$ with $B\in C^0\left(\R^3,\R^3\right)$ such that $\mathrm{div}B=0$ and satisfying the assumptions \eqref{eq:main-thm-2-hp}. By \cite[Lemma~5]{GraLewSer-2018-JMPA}, let $A$ be the unique potential in $\Cgaugesobolev$ such that $B=\mathrm{curl}A$, and let $A_\varepsilon$ be the potential associated to $B_\varepsilon$. We are interested in studying the asymptotic behaviour of the free energy functional $\mathcal{F}_\mathrm{PV}\left(A_\varepsilon,\beta\right)$ of Dirac's vacuum as $\varepsilon\xrightarrow{}0$, that is in the limit of slowly varying purely magnetic fields.\par
Similarly to what has been done in \cref{subsec:integrable-magnetic-potentials}, we rewrite Formula~\eqref{eq:tanh-square} as follows:
\begin{equation}\label{eq:tanh-square-bis}
    x\tanh{x}=\sum_{k\in\Z}\frac{4x^2}{\left(2k-1\right)^2\pi^2+4x^2}=\sum_{k\in\Z}\left(1-\frac{\left(2k-1\right)^2\pi^2}{\left(2k-1\right)^2\pi^2+4x^2}\right),
\end{equation}
in order to express the free energy functional $\mathcal{F}_\mathrm{PV}$ of Dirac's vacuum as
\begin{multline*}
    \mathcal{F}_\mathrm{PV}\left(A,\beta\right)=\frac{1}{\beta}\int_0^\beta\tr\left[ T_A\left(b\right)\right]\diff b\\
    =\frac{1}{\beta}\int_0^\beta \sum_{k\in\Z}\tr\left[\sum_{j=0}^2 c_j\left(\frac{1}{\left(D^A_{m_j}\right)^2+\omega\left(k,b\right)^2}-\frac{1}{\left(D^0_{m_j}\right)^2+\omega\left(k,b\right)^2}\right)\right]\\
    \times\omega\left(k,b\right)^2\frac{db}{b}.
\end{multline*}
Notice that the series is convergent in the operator norm thanks to the following estimate:
\[
\norm{\sum_{j=0}^2 c_j\left(\frac{1}{\left(D^A_{m_j}\right)^2+\omega\left(k,\beta\right)^2}-\frac{1}{\left(D^0_{m_j}\right)^2+\omega\left(k,\beta\right)^2}\right)}\leq \frac{C}{\left(m^2+\omega\left(k,\beta\right)^2\right)^3}.
\]
In this section, we prefer working with Formula~\eqref{eq:tanh-square-bis} rather than \eqref{eq:tanh-linear} in order to make the square of the Dirac operator appear, which is positive definite and allows us to rewrite $T_A\left(b\right)$ in a block-diagonal form in the two-spinor basis:
\[
T_A\left(b\right)=\frac{1}{b}\sum_{k\in\Z}\omega\left(k,b\right)^2\sum_{j=0}^2 c_j\left(\frac{1}{\mathcal{P}_A+m_j^2+\omega\left(k,b\right)^2}-\frac{1}{\mathcal{P}_0+m_j^2+\omega\left(k,b\right)^2}\right)\otimes \mathds{1}_{\C^2},
\]
where $\mathcal{P}_A$ is the Pauli operator defined as $\mathcal{P}_A=\left(\sigma\cdot\left(-i\nabla+A\right)\right)^2=\left(-i\nabla+A\right)^2-\sigma\cdot B$ such that
\[
\abs{D^A_m}^2=\begin{bmatrix}
\mathcal{P}_A+m^2 & 0\\
0 & \mathcal{P}_A+m^2
\end{bmatrix}=\left(\mathcal{P}_A+m^2\right)\otimes\mathds{1}_{\C^2}.
\]
By \cref{stat:main-thm-1}, the operator $T_A\left(b\right)$ is trace class when $A\in L^1\left(\R^3,\R^3\right)\cap H^1\left(\R^3,\R^3\right)$, but in this section we simply have $B\in L^1\left(\R^3\right)\cap L^\infty\left(\R^3\right)$, which does not imply the associated potential $A$ to be integrable. Thus, $T_A\left(b\right)$ is not necessarily trace class in our framework. Fortunately, the non-trace-class part can be easily isolated and is linear in $A$. Indeed, by adjusting \cite[Proposition~6]{GraLewSer-2018-JMPA} to the non-zero temperature case by means of the previously mentioned replacements, we know that, under our assumptions on $B$,
\begin{multline}\label{eq:free-energy-minus-non-trace-class-term}
    \mathcal{F}_\mathrm{PV}\left(A,\beta\right)=\frac{1}{\beta}\int_0^\beta \sum_{k\in\Z}\tr\left[\sum_{j=0}^2 c_j\left(\frac{1}{\mathcal{P}_A+m_j^2+\omega\left(k,b\right)^2}-\frac{1}{\mathcal{P}_0+m_j^2+\omega\left(k,b\right)^2}\right.\right.\\
    \left.\left.-\frac{1}{\mathcal{P}_0+m_j^2+\omega\left(k,b\right)^2}\left(p\cdot A+A\cdot p\right)\frac{1}{\mathcal{P}_0+m_j^2+\omega\left(k,b\right)^2}\right)\right]\omega\left(k,b\right)^2\frac{\diff b}{b}.
\end{multline}
As done in \cite[Step~3]{GraLewSer-2018-JMPA}, we now introduce a Gaussian function $G_\rho$ given by
\[
G_\rho\left(x\right)=\left(\pi\rho\right)^{-\frac{3}{2}}e^{-\frac{\abs{x}^2}{\rho^2}},
\]
and such that
\[
\int_{\R^3}G_\rho\left(x-y\right)^2\diff y=1,\quad\forall x\in\R^3,
\]
which will play the role of a continuous partition of unity and allow us to localise the free energy functional in order to work on sets of fixed size where the slowly varying magnetic field is almost constant. We refer the reader to \cite[Lemma~7]{GraLewSer-2018-JMPA} for some properties about this Gaussian family. Now, by localising the operator inside the parenthesis by means of the functions $G_\rho$, the functional \eqref{eq:free-energy-minus-non-trace-class-term} can be rewritten as
\[
\mathcal{F}_\mathrm{PV}\left(A_\varepsilon,\beta\right)=\frac{2\varepsilon^{-3}}{\beta}\int_0^\beta\int_{\R^3}\sum_{k\in\Z}f_{\omega\left(k,b\right)}\left(A_{\varepsilon,y}\right)\omega\left(k,b\right)^2\diff y\frac{\diff b}{b},
\]
where
\[
A_{\varepsilon,y}\left(x\right)=\varepsilon^{-1}A\left(y+\varepsilon x\right)
\]
and
\begin{multline*}
    f_\omega\left(A\right)=\tr_{L^2\left(\R^3,\C^2\right)}\left[G_\rho\sum_{j=0}^2 c_j\left(\frac{1}{\mathcal{P}_A+m_j^2+\omega^2}-\frac{1}{\mathcal{P}_0+m_j^2+\omega^2}\right.\right.\\
    \left.\left.-\frac{1}{\mathcal{P}_0+m_j^2+\omega^2}\left(p\cdot A+A\cdot p\right)\frac{1}{\mathcal{P}_0+m_j^2+\omega^2}\right)G_\rho\right].
\end{multline*}
Thanks to the localisation procedure, the term containing $p\cdot A+A\cdot p$, which was subtracted because of its non-trace-class nature, becomes trace class with vanishing trace \cite[Lemma~8]{GraLewSer-2018-JMPA}. As a consequence, $f_\omega$ can be simplified as follows:
\[
f_\omega\left(A_{\varepsilon,y}\right)=\tr_{L^2\left(\R^3,\C^2\right)}\left[G_\rho\sum_{j=0}^2 c_j\left(\frac{1}{\mathcal{P}_{A_{\varepsilon,y}}+m_j^2+\omega^2}-\frac{1}{\mathcal{P}_0+m_j^2+\omega^2}\right)G_\rho\right].
\]
Now, the idea is to replace the potential $A_{\varepsilon,y}$ in the formula above by the potential $B\left(y\right)\times x/2$ of the constant field $B\left(y\right)$, up to a small error. In order to do so, we first need to extend the definition of $f_\omega$ to potentials with a linear growth at infinity. This is doable thanks to \cite[Proposition~9]{GraLewSer-2018-JMPA}. In particular, the functional $f_\omega$ is gauge invariant, meaning that
\[
f_\omega\left(A\right)=f_\omega\left(A+\nabla\theta\right),\quad\text{for any }\theta\in C^2\left(\R^3,\R\right).
\]
Then, applying \cite[Formula~(55)]{GraLewSer-2018-JMPA} to $A_{\varepsilon,y}$ gives
\begin{equation}\label{eq:magnetic-potential-decomposition}
    \frac{A\left(y+\varepsilon x\right)}{\varepsilon}=\nabla_x\left(x\cdot\int_0^1\frac{A\left(y+t\varepsilon x\right)}{\varepsilon}\diff t\right)+B\left(y\right)\times\frac{x}{2}+\varepsilon R_{\varepsilon,y}\left(x\right),
\end{equation}
where
\[
R_{\varepsilon,y}\left(x\right)=x\times\int_0^1\frac{B\left(y\right)-B\left(y+t\varepsilon x\right)}{\varepsilon}t\diff t.
\]
The main idea now is that the gradient term in \eqref{eq:magnetic-potential-decomposition} can be removed in the argument of the functional $f_\omega$ thanks to its gauge invariance property. The remaining potential does not satisfy the Coulomb gauge anymore, but rather the Poincaré (also called multipolar or Fock-Schwinger) gauge,
\[
x\cdot A\left(x+y\right)=0.
\]
Finally, the free energy functional $\mathcal{F}_\mathrm{PV}$ of Dirac's vacuum can be rewritten as
\begin{multline}\label{eq:T-EH-final-formula}
    \varepsilon^3\mathcal{F}_\mathrm{PV}\left(A_\varepsilon,\beta\right)=\frac{2}{\beta}\int_0^\beta\int_{\R^3}\sum_{k\in\Z}\omega\left(k,b\right)^2 f_{\omega\left(k,b\right)}\left(A_{\varepsilon,y}\right)\diff y\frac{\diff b}{b}\\
    =\frac{2}{\beta}\int_0^\beta\int_{\R^3}\sum_{k\in\Z}\omega\left(k,b\right)^2 f_{\omega\left(k,b\right)}\left(B\left(y\right)\times\frac{\cdot}{2} +\varepsilon R_{\varepsilon,y}\right)\diff y\frac{\diff b}{b}\\
    =\int_{\R^3}\left[\frac{2}{\beta}\int_0^\beta\sum_{k\in\Z}\omega\left(k,b\right)^2 f_{\omega\left(k,b\right)}\left(B\left(y\right)\times\frac{\cdot}{2}\right)\frac{\diff b}{b}\right]\diff y\\
    +\frac{2}{\beta}\int_0^\beta\int_{\R^3}\sum_{k\in\Z}\omega\left(k,b\right)^2\left[f_{\omega\left(k,b\right)}\left(B\left(y\right)\times\frac{\cdot}{2}+\varepsilon R_{\varepsilon,y}\right)-f_{\omega\left(k,b\right)}\left(B\left(y\right)\times\frac{\cdot}{2}\right)\right]\diff y\frac{\diff b}{b}.
\end{multline}
Thus, it only remains to study the two terms in \cref{eq:T-EH-final-formula}. This is the content of the next two lemmas. We start by computing an explicit formula for the first term:
\begin{lem}\label{stat:T-EH-formula}
    Under the assumptions of \cref{stat:main-thm-2},
    \[
    \frac{2}{\beta}\int_0^\beta\sum_{k\in\Z}\omega\left(k,b\right)^2 f_{\omega\left(k,b\right)}\left(B\left(y\right)\times\frac{\cdot}{2}\right)\frac{\diff b}{b}=f_\mathrm{PV}^0\left(\abs{B\left(y\right)}\right)+f_\mathrm{PV}^T\left(\abs{B\left(y\right)},\beta\right),
    \]
    where
    \[
    f_\mathrm{PV}^0\left(a\right)=\frac{1}{8\pi^2}\int_0^\infty\left(\sum_{j=0}^2c_je^{-sm_j^2}\right)\left(sa\coth\left(sa\right)-1\right)\frac{\diff s}{s^3}
    \]
    and
    \begin{equation}\label{eq:T-f-PV}
        f^T_\mathrm{PV}\left(a,\beta\right)=\frac{1}{4\pi^2}\int_0^\infty\sum_{j=0}^2 c_j e^{-sm_j^2}\left(s\abs{B}\coth\left(s\abs{B}\right)-1\right)\sum_{n=1}^\infty \left(-1\right)^n e^{-\frac{\beta^2 n^2}{4s}}\frac{\diff s}{s^3}.
    \end{equation}
\end{lem}
\proof
By \cite[Proposition~16]{GraLewSer-2018-JMPA}, we know that, if $B$ is constant,
\[
f_\omega\left(B\times\frac{x}{2}\right)=\frac{1}{4\pi^{3/2}}\int_0^\infty e^{-s\omega^2}\left(\sum_{j=0}^2 c_j e^{-sm_j^2}\right)\left(s\abs{B}\coth\left(s\abs{B}\right)-1\right)\frac{\diff s}{s^{3/2}}.
\]
Therefore,
\begin{multline*}
    \frac{2}{\beta}\int_0^\beta\sum_{k\in\Z}\omega\left(k,b\right)^2 f_{\omega\left(k,b\right)}\left(B\times\frac{\cdot}{2}\right)\frac{\diff b}{b}\\
    =\frac{1}{2\beta\pi^{3/2}}\int_0^\beta\int_0^\infty\left(\sum_{j=0}^2 c_j e^{-sm_j^2}\right)\left(s\abs{B}\coth\left(s\abs{B}\right)-1\right)\sum_{k\in\Z}\omega\left(k,b\right)^2 e^{-s\omega\left(k,b\right)^2}\frac{\diff s}{s^{3/2}}\frac{\diff b}{b}\\
    =-\frac{1}{2\beta\pi^{3/2}}\int_0^\beta\int_0^\infty\left(\sum_{j=0}^2 c_j e^{-sm_j^2}\right)\left(s\abs{B}\coth\left(s\abs{B}\right)-1\right)\frac{\diff}{\diff s}\theta_2\left(0,\frac{4\pi i s}{b^2}\right)\frac{\diff s}{s^{3/2}}\frac{\diff b}{b}\\
    =\frac{1}{2\beta\pi^{3/2}}\int_0^\beta\int_0^\infty\frac{\diff}{\diff s}\left[\frac{1}{s^{3/2}}\left(\sum_{j=0}^2 c_j e^{-sm_j^2}\right)\left(s\abs{B}\coth\left(s\abs{B}\right)-1\right)\right]\theta_2\left(0,\frac{4\pi i s}{b^2}\right)\diff s\frac{\diff b}{b},
\end{multline*}
where $\theta_2$ is defined by \eqref{eq:jacobi-theta-2} and $B$ denotes $B\left(y\right)$ for fixed $y$. As done in the proof of \cref{stat:explicit-second-order}, we apply again Dittrich's method \cite{Dit-1979-PRD} in order to isolate the non-zero temperature contribution from the null temperature term. Thus, thanks to \cref{eq:theta-function-poisson}, we can rewrite the equation above as follows:
\[
\frac{2}{\beta}\int_0^\beta\sum_{k\in\Z}\omega\left(k,b\right)^2 f_{\omega\left(k,b\right)}\left(B\times\frac{\cdot}{2}\right)\frac{\diff b}{b}=f^0_{\mathrm{PV}}\left(\abs{B}\right)+f^T_{\mathrm{PV}}\left(\abs{B},\beta\right),
\]
where
\[
f^0_\mathrm{PV}\left(\abs{B}\right)=\frac{1}{4\pi^2}\int_0^\infty\frac{\diff}{\diff s}\left[\frac{1}{s^{3/2}}\left(\sum_{j=0}^2 c_j e^{-sm_j^2}\right)\left(s\abs{B}\coth\left(s\abs{B}\right)-1\right)\right]\frac{1}{s^{1/2}}\diff s
\]
and
\begin{multline*}
    f^T_\mathrm{PV}\left(\abs{B},\beta\right)=\frac{1}{2\beta\pi^2}\int_0^\beta\int_0^\infty\frac{\diff}{\diff s}\left[\frac{1}{s^{3/2}}\left(\sum_{j=0}^2 c_j e^{-sm_j^2}\right)\left(s\abs{B}\coth\left(s\abs{B}\right)-1\right)\right]\\
    \times\frac{1}{s^{1/2}}\sum_{n=1}^\infty \left(-1\right)^n e^{-\frac{b^2n^2}{4s}}\diff s\diff b.
\end{multline*}
Let us now focus on $f^0_\mathrm{PV}$. By computing the derivative of the product and integrating by parts once more one of the resulting terms, we finally get
\[
f^0_\mathrm{PV}\left(\abs{B}\right)=\frac{1}{8\pi^2}\int_0^\infty\left(\sum_{j=0}^2 c_j e^{-sm_j^2}\right)\left(s\abs{B}\coth\left(s\abs{B}\right)-1\right)\frac{\diff s}{s^3},
\]
which is the well known Euler-Heisenberg energy function at null temperature \cite{GraLewSer-2018-JMPA}. Then, let us focus on $f^T_\mathrm{PV}$. Again, similarly to what we did for the zero temperature term, we get
\begin{multline*}
    f^T_\mathrm{PV}\left(\abs{B},\beta\right)=\frac{1}{4\beta\pi^2}\int_0^\beta\int_0^\infty\sum_{j=0}^2 c_j e^{-sm_j^2}\left(s\abs{B}\coth\left(s\abs{B}\right)-1\right)\sum_{n=1}^\infty\left(-1\right)^n e^{-\frac{b^2 n^2}{4s}}\frac{\diff s}{s^3}\diff b\\
    -\frac{1}{4\beta\pi^2}\int_0^\beta\int_0^\infty\sum_{j=0}^2 c_j e^{-sm_j^2}\left(s\abs{B}\coth\left(s\abs{B}\right)-1\right)\sum_{n=1}^\infty\left(-1\right)^n \frac{b^2 n^2}{4s} e^{-\frac{b^2 n^2}{2s}}\frac{\diff s}{s^3}\diff b.
\end{multline*}
Noticing that
\[
\frac{1}{\beta}\int_0^\beta\left(1-\frac{b^2n^2}{2s}\right)e^{-\frac{b^2n^2}{4s}}\diff b=e^{-\frac{\beta^2n^2}{4s}},
\]
we get \cref{eq:T-f-PV} and this allows us to conclude the proof.
\endproof
Finally, it only remains to give an estimate of the remainder in the second line of \eqref{eq:T-EH-final-formula}:
\begin{lem}\label{stat:T-EH-remainder}
    Under the assumptions of \cref{stat:main-thm-2}, we have, for every $0<\varepsilon\leq 1$,
    \[
    \int_{\R^3}\abs{f_{\omega\left(k,b\right)}\left(B\left(y\right)\times\frac{\cdot}{2}+\varepsilon R_{\varepsilon,y}\right)-f_{\omega\left(k,b\right)}\left(B\left(y\right)\times\frac{\cdot}{2}\right)}\diff y\leq C\frac{\varepsilon}{\left(m^2+\omega\left(k,b\right)^2\right)^2},
    \]
    where $C$ is a constant depending only on $B$, $\rho$ and the $c_j$'s and $m_j$'s.
\end{lem}
\proof
The proof is the same as in \cite[Proposition~17]{GraLewSer-2018-JMPA}.
\endproof
Formula \eqref{eq:T-EH-final-formula} together with \cref{stat:T-EH-formula} and \cref{stat:T-EH-remainder} allows us to conclude the proof.\qed

\appendix
\section{The non-zero temperature regularisation in the ultraviolet regime}\label{sec:appendix-A}
In this appendix, we consider the difference between the relative free energy and the relative energy of each virtual particle and prove that this difference is finite and tends to zero as the mass of the virtual particle tends to infinity. Notice that by “relative” we mean the energy (free or not) with $A$ minus the energy (free or not) without $A$. This proves that the non-zero temperature regularisation converges to the null-temperature one in the ultraviolet regime. Therefore, the same renormalisation formula at zero temperature can be applied here to remove the unphysical regularisation parameters, as explained in \cref{rmk:zero-temperature-renormalisation-formula}.\par
Formally, for a fixed $\beta>0$, we consider the following functional
\begin{equation*}
    \widetilde{\Xi}\left(m_j,\beta,A\right)=\frac{1}{\beta}\mathrm{tr}\left[\log\left(2\cosh\frac{\beta D^0_{m_j}}{2}\right)-\log\left(2\cosh\frac{\beta D^A_{m_j}}{2}\right)\right]-\frac{1}{2}\mathrm{tr}\left[\vert D^0_{m_j} \vert -\vert D^A_{m_j}\vert \right]
\end{equation*}
representing the difference between the relative free energy and the relative energy of the virtual particle of mass $m_j>0$. In order to make the notations lighter, $m_j$ will be simply denoted by $m$ from now on. As done in \cref{eq:integral-free-energy}, $\widetilde{\Xi}$ can be rewritten as
\[
\widetilde{\Xi}\left(m,\beta,A\right)=\frac{1}{2\beta}\int_0^\beta\mathrm{tr}\left[D^0_m\tanh\frac{bD^0_m}{2}-D^A_m\tanh\frac{bD^A_m}{2}\right]db-\frac{1}{2}\mathrm{tr}\left[\vert D^0_m \vert -\vert D^A_m\vert \right].
\]
Of course, none of the two operators inside the trace is trace class. Therefore, in order to work with well defined quantities, it is more convenient to consider the functional in the following form
\[
\Xi\left(m,\beta,A\right)=\frac{1}{2\beta}\int_0^\beta\mathrm{tr}\left[D^0_m\tanh\frac{bD^0_m}{2}-D^A_m\tanh\frac{bD^A_m}{2}-\vert D^0_m \vert +\vert D^A_m\vert \right]db.
\]
The main content of this appendix is the following result. Notice that we require $B=\mathrm{curl}A\in L^1\left(\R^3,\R^3\right)$ in addition to assumptions of \cref{stat:main-thm-1}. This hypothesis, which is not restrictive both from a mathematical and from a physical point of view, is purely technical, arising from the proof below, and could probably be removed.
\begin{thm}
    Let $\beta>0$. Let $A\in L^1\left(\R^3,\R^3\right)\cap\Cgaugesobolev$ such that $B=\mathrm{curl}A\in L^1\left(\R^3,\R^3\right)$. Then, the functional $\Xi$ is well defined and
    \[
    \lim_{m\to+\infty} \Xi\left(m,\beta,A\right)=0.
    \]
\end{thm}
\proof
    Our proof reduces to finding some trace-class norm estimates of
    \begin{equation*}
        \xi\left(m,b,A\right)=D^0_m\tanh\frac{bD^0_m}{2}-D^A_m\tanh\frac{bD^A_m}{2}-\vert D^0_m \vert +\vert D^A_m\vert
    \end{equation*}
    that go to zero as $m$ goes to infinity. By means of the arguments in \cref{subsec:integrable-magnetic-potentials} and in \cite{GraHaiLewSer-2013-ARMA}, $\xi$ can be rewritten as
    \begin{multline*}
        \xi\left(m,b,A\right)=\frac{1}{b}\sum_{k\in\mathbb{Z}}\left(\frac{i\omega\left(k,b\right)}{D^A_m+i\omega\left(k,b\right)}-\frac{i\omega\left(k,b\right)}{D^A_m-i\omega\left(k,b\right)}-\frac{i\omega\left(k,b\right)}{D^0_m+i\omega\left(k,b\right)}+\frac{i\omega\left(k,b\right)}{D^0_m-i\omega\left(k,b\right)}\right)\\
        -\frac{1}{2\pi}\int_\mathbb{R}\left(\frac{i\omega}{D^A_m+i\omega}-\frac{i\omega}{D^A_m-i\omega}-\frac{i\omega}{D^0_m+i\omega}+\frac{i\omega}{D^0_m-i\omega}\right)d\omega.
    \end{multline*}
    A simple change of variable (recall that $\omega\left(k,b\right)=\left(2k-1\right)\pi/b$) yields
    \begin{multline*}
        \xi\left(m,b,A\right)=\frac{1}{b}\sum_{k\in\mathbb{Z}}\left(\frac{i\omega\left(k,b\right)}{D^A_m+i\omega\left(k,b\right)}-\frac{i\omega\left(k,b\right)}{D^A_m-i\omega\left(k,b\right)}-\frac{i\omega\left(k,b\right)}{D^0_m+i\omega\left(k,b\right)}+\frac{i\omega\left(k,b\right)}{D^0_m-i\omega\left(k,b\right)}\right)\\
        -\frac{1}{b}\int_\mathbb{R}\left(\frac{i\omega\left(k,b\right)}{D^A_m+i\omega\left(k,b\right)}-\frac{i\omega\left(k,b\right)}{D^A_m-i\omega\left(k,b\right)}-\frac{i\omega\left(k,b\right)}{D^0_m+i\omega\left(k,b\right)}+\frac{i\omega\left(k,b\right)}{D^0_m-i\omega\left(k,b\right)}\right)d\omega
    \end{multline*}
    and shows that the operator taken into account is given by the difference between an integral and its discrete series. If we now denote
    \[
    \Theta\left(k,b\right)=\frac{i\omega\left(k,b\right)}{D^A_m+i\omega\left(k,b\right)}-\frac{i\omega\left(k,b\right)}{D^A_m-i\omega\left(k,b\right)}-\frac{i\omega\left(k,b\right)}{D^0_m+i\omega\left(k,b\right)}+\frac{i\omega\left(k,b\right)}{D^0_m-i\omega\left(k,b\right)},
    \]
    the Euler-Maclaurin formula allows us to write
    \begin{align*}
        \xi\left(m,b,A\right)&=\sum_{k\in\mathbb{Z}}\Theta\left(k,b\right)-\int_\mathbb{R}\Theta\left(k,b\right)dk\\
        &=\underbrace{\frac{\Theta\left(+\infty,b\right)+\Theta\left(-\infty,b\right)}{2}}_{=0}+\int_\mathbb{R}\Theta'\left(k,b\right)P_1\left(k\right)dk,
    \end{align*}
    where $P_1\left(x\right)=B_1\left(x-\lfloor x \rfloor\right)$, $B_1$ being a Bernoulli function. Of course, the last formula can be rigorously justified by first working on the interval $\left[-M,M\right]$ and then taking the limit $M\to\infty$. Finally, it suffices to find a trace-class norm estimate for the operator $\Theta'\left(k,b\right)$, integrable with respect to $k$ on $\mathbb{R}$ and which goes to zero as $m$ goes to infinity. This operator is equal to
    \begin{multline*}
       \Theta'\left(k,b\right)=\frac{2\pi i}{b}\left[\frac{D^A_m}{\left(D^A_m+i\omega\left(k,b\right)\right)^2}-\frac{D^A_m}{\left(D^A_m-i\omega\left(k,b\right)\right)^2}\right.\\
       \left.-\frac{D^0_m}{\left(D^0_m+i\omega\left(k,b\right)\right)^2}+\frac{D^0_m}{\left(D^0_m-i\omega\left(k,b\right)\right)^2}\right] 
    \end{multline*}
    which can be easily rewritten as
    \[
    \Theta'\left(k,b\right)=\frac{4\pi\omega\left(k,b\right)}{b}\left[\frac{\left(D^A_m\right)^2}{\left(\left(D^A_m\right)^2+\omega\left(k,b\right)^2\right)^2}-\frac{\left(D^0_m\right)^2}{\left(\left(D^0_m\right)^2+\omega\left(k,b\right)^2\right)^2}\right].
    \]
    Since $D^A_m=\boldsymbol{\alpha}\cdot\left(-i\nabla+A\right)+m\beta$, one can show that
    \[
    \left(D^A_m\right)^2=\left(-\Delta+m^2+\boldsymbol{\sigma}\cdot B+\vert A\vert^2\right)\otimes\mathds{1}_{\mathbb{C}^2}.
    \]
    Consequently,
    \begin{multline*}
        \frac{\left(D^A_m\right)^2}{\left(\left(D^A_m\right)^2+\omega\left(k,b\right)^2\right)^2}-\frac{\left(D^0_m\right)^2}{\left(\left(D^0_m\right)^2+\omega\left(k,b\right)^2\right)^2}=\\
        \left(\left(-\Delta+m^2\right)\otimes\mathds{1}_{\mathbb{C}^2}\right)\left[\frac{1}{\left(\left(D^A_m\right)^2+\omega\left(k,b\right)^2\right)^2}-\frac{1}{\left(\left(D^0_m\right)^2+\omega\left(k,b\right)^2\right)^2}\right]\\
        +\left(\left(\boldsymbol{\sigma}\cdot B+\vert A\vert^2\right)\otimes\mathds{1}_{\mathbb{C}^2}\right)\frac{1}{\left(\left(D^A_m\right)^2+\omega\left(k,b\right)^2\right)^2}.
    \end{multline*}
    Using that
    \begin{multline*}
        \left(D^A_m\right)^2+\omega\left(k,b\right)^2=\\
        \left(1+\left(\boldsymbol{\sigma}\cdot B+\vert A\vert^2\right)\left(-\Delta+m^2+\omega\left(k,b\right)^2\right)^{-1}\right)\left(-\Delta+m^2+\omega\left(k,b\right)^2\right)\otimes\mathds{1}_{\mathbb{C}^2},
    \end{multline*}
    we can rewrite
    \begin{multline*}
        \frac{\left(D^A_m\right)^2}{\left(\left(D^A_m\right)^2+\omega\left(k,b\right)^2\right)^2}-\frac{\left(D^0_m\right)^2}{\left(\left(D^0_m\right)^2+\omega\left(k,b\right)^2\right)^2}=\\
        \frac{-\Delta+m^2}{\left(-\Delta+m^2+\omega\left(k,b\right)^2\right)^2}\left[\left(\underbrace{1+\left(\boldsymbol{\sigma}\cdot B+\vert A\vert^2\right)\left(-\Delta+m^2+\omega\left(k,b\right)^2\right)^{-1}}_{\mathcal{O}}\right)^{-2}-1\right]\otimes\mathds{1}_{\mathbb{C}^2}\\
        +\frac{\boldsymbol{\sigma}\cdot B+\vert A\vert^2}{\left(-\Delta+m^2+\omega\left(k,b\right)^2\right)^2}\left[1+\left(\boldsymbol{\sigma}\cdot B+\vert A\vert^2\right)\left(-\Delta+m^2+\omega\left(k,b\right)^2\right)^{-1}\right]^{-2}\otimes \mathds{1}_{\mathbb{C}^2}\\
        \eqqcolon \mathcal{A}+\mathcal{B}.
    \end{multline*}
    Notice that $\mathcal{O}$ is invertible provided that $m$ is large enough (and this is not a problem since we will be ultimately interested in the large-mass limit).\\
    Now,
    \begin{multline*}
        \left(1+\left(\boldsymbol{\sigma}\cdot B+\vert A\vert^2\right)\left(-\Delta+m^2+\omega\left(k,b\right)^2\right)^{-1}\right)^{-2}-1\\
        =\left(\sum_{k=0}^\infty \left(-\left(\boldsymbol{\sigma}\cdot B+\vert A\vert^2\right)\left(-\Delta+m^2+\omega\left(k,b\right)^2\right)^{-1}\right)^k\right)^2-1\\
        =\sum_{k=1}^\infty\left(-\left(\boldsymbol{\sigma}\cdot B+\vert A\vert^2\right)\left(-\Delta+m^2+\omega\left(k,b\right)^2\right)^{-1}\right)^k\\
        \times\left[2+\sum_{k=1}^\infty \left(-\left(\boldsymbol{\sigma}\cdot B+\vert A\vert^2\right)\left(-\Delta+m^2+\omega\left(k,b\right)^2\right)^{-1}\right)^k\right]\\
        =-\left(\boldsymbol{\sigma}\cdot B+\vert A\vert^2\right)\left(-\Delta+m^2+\omega\left(k,b\right)^2\right)^{-1}\mathcal{L},
    \end{multline*}
    where $\mathcal{L}$ is a bounded operator given by convergent Neumann series (once again, for $m$ large enough), which allows us to rewrite $\mathcal{A}$ as
    \begin{multline*}
        \mathcal{A}=-\left[\frac{-\Delta+m^2}{\left(-\Delta+m^2+\omega\left(k,b\right)^2\right)^2}\big\vert\boldsymbol{\sigma}\cdot B+\vert A\vert^2\big\vert^{1/2}\mathrm{sign}\left(\boldsymbol{\sigma}\cdot B+\vert A\vert^2\right)\right]\\
        \times\frac{\big\vert\boldsymbol{\sigma}\cdot B+\vert A\vert^2\big\vert^{1/2}}{-\Delta+m^2+\omega\left(k,b\right)^2}\mathcal{L}\otimes\mathds{1}_{\mathbb{C}^2}\\
        \eqqcolon -\mathcal{A}_1 \mathcal{A}_2\otimes\mathds{1}_{\mathbb{C}^2}.
    \end{multline*}
    Thus, in order to get a trace-class norm estimate for the operator $\mathcal{A}$, it suffices to separately estimate the Hilbert-Schmidt norm of $\mathcal{A}_1$ and $\mathcal{A}_2$. Recalling that
    \[
    \frac{-\Delta+m^2}{-\Delta+m^2+\omega\left(k,b\right)^2},\quad\mathrm{sign}\left(\boldsymbol{\sigma}\cdot B+\vert A\vert^2\right)\quad\text{and}\quad\mathcal{L}
    \]
    are bounded operators (uniformly in $m$), only the $\mathfrak{S}_2$-norm of $\left(-\Delta+m^2+\omega\left(k,b\right)\right)^{-1}\big\vert\boldsymbol{\sigma}\cdot B+\vert A\vert^2\big\vert^{1/2}$ needs to be computed. We have\footnote{Recall that $\left(-\Delta+m^2+\omega^2\right)^{-1}$ has an integral kernel given by $\frac{e^{-\sqrt{m^2+\omega^2}\vert x-y\vert}}{4\pi\vert x-y\vert}$.}
    \begin{multline*}
        \left\lVert \left(-\Delta+m^2+\omega\left(k,b\right)\right)^{-1}\left\lvert\boldsymbol{\sigma}\cdot B+\left\lvert A\right\rvert^2\right\rvert^{1/2}\right\rVert_{\mathfrak{S}_2}^2\\
        =\iint \left\lvert\frac{e^{-\sqrt{m^2+\omega\left(k,b\right)^2}\left\lvert x-y\right\rvert}}{4\pi\left\lvert x-y\right\rvert}\left\lvert\boldsymbol{\sigma}\cdot B\left(y\right)+\left\lvert A\left(y\right)\right\rvert^2\right\rvert^{1/2}\right\rvert^2 dxdy\\
        \leq \iint \frac{e^{-2\sqrt{m^2+\omega\left(k,b\right)^2}\left\lvert x-y\right\rvert}}{8\pi^2\left\lvert x-y\right\rvert^2}\left(\left\lvert B\left(y\right)\right\rvert+\left\lvert A\left(y\right)\right\rvert^2\right) dxdy\\
        =\left\lVert\mathcal{G}\ast\left\lvert B\right\rvert\right\rVert_{L^1}+\left\lVert\mathcal{G}\ast\left\lvert A\right\rvert^2\right\rVert_{L^1},
    \end{multline*}
    where
    \[
    \mathcal{G}\left(x\right)=\frac{e^{-2\sqrt{m^2+\omega\left(k,b\right)^2}\vert x\vert}}{8\pi^2\vert x\vert^2}.
    \]
    By Young's inequality,
    \[
    \left\lVert\mathcal{G}\ast\left\lvert A\right\rvert^2\right\rVert_{L^1}\leq \left\lVert \mathcal{G}\right\rVert_{L^1}\left\lVert A\right\rVert_{L^2}<\infty
    \]
    and
    \[
    \left\lVert\mathcal{G}\ast\left\lvert B\right\rvert\right\rVert_{L^1}\leq \left\lVert \mathcal{G}\right\rVert_{L^1}\left\lVert B\right\rVert_{L^1}<\infty.
    \]
    Notice that we need here the additional assumption that $B\in L^1\left(\R^3,\R^3\right)$. Moreover, by a simple change of variable, the $L^1$-norm of $\mathcal{G}$ is proved to be integrable with respect to $k$ over the whole real line and to go to zero as $m$ goes to infinity, as expected. Indeed,
    \[
    \left\lVert\mathcal{G}\right\rVert_{L^1}\leq\frac{C}{8\pi^2}\frac{1}{\left(m^2+\omega\left(k,b\right)^2\right)^{5/2}}\xrightarrow{m\to+\infty}0.
    \]
    In short, we proved $\mathcal{A}$ to be a trace-class operator whose trace-class norm is integrable with respect to $k$ over $\mathbb{R}$ and converges to zero in the large-mass limit.\\
    A similar argument works also for
    \begin{multline*}
        \mathcal{B}=\left(-\Delta+m^2+\omega\left(k,b\right)^2\right)^{-2}\left\lvert\boldsymbol{\sigma}\cdot B+\vert A\vert^2\right\rvert^{1/2}\mathrm{sign}\left(\boldsymbol{\sigma}\cdot B+\vert A\vert^2\right)\\
        \times\left\lvert\boldsymbol{\sigma}\cdot B+\vert A\vert^2\right\rvert^{1/2}\left[1+\left(\boldsymbol{\sigma}\cdot B+\vert A\vert^2\right)\left(-\Delta+m^2+\omega\left(k,b\right)^2\right)^{-1}\right]^{-2}\otimes\mathds{1}_{\mathbb{C}^2}\\
        \eqqcolon\mathcal{B}_1\mathcal{B}_2\otimes\mathds{1}_{\mathbb{C}^2},
    \end{multline*}
    by treating $\mathcal{B}_1$ and $\mathcal{B}_2$ separately. This concludes the proof.
\endproof

\addtocontents{toc}{\protect\setcounter{tocdepth}{0}} 

\section*{Acknowledgements}
The author would like to thank \'Eric Séré for useful remarks and fruitful discussions and William Borrelli for the time spent together in finding the right argument for the proof in \cref{sec:appendix-A}. This research project has received funding from the European Union's Horizon 2020 research and innovation programme under the Marie Skłodowska-Curie grant agreement N°945332. The author has also been funded by the Italian Ministry of University and Research (MUR) and Next Generation EU under the PRIN 2022
Project [Prot.2022AKRC5P\_\emph{Interacting quantum systems: topological phenomena and effective theories}].

\section*{Conflict of interest}
The author declares no conflict of interest.

\section*{Data availability statement}
Data sharing is not applicable to this article as it has no associated data. 

\addtocontents{toc}{\protect\setcounter{tocdepth}{2}} 

\bibliographystyle{siam} 
\small{\bibliography{bibliography}}

\begin{thebibliography}{10}

\bibitem{AbrSte-1964-book}
{\sc M.~Abramowitz and I.~A. Stegun}, {\em Handbook of mathematical functions with formulas, graphs and mathematical tables}, United States Department of Commerce, National Bureau of Standards, 1964.

\bibitem{BacLieSol-1994-JSP}
{\sc V.~Bach, E.~H. Lieb, and J.~P. Solovej}, {\em Generalized {H}artree-{F}ock theory and the {H}ubbard model}, Journal of Statistical Physics, 76 (1994), pp.~3--89.

\bibitem{BarHar-2001-TAJ}
{\sc M.~G. Baring and A.~K. Harding}, {\em Photon splitting and pair creation in highly magnetized pulsars}, The Astrophysical Journal, 547 (2001), pp.~929--948.

\bibitem{Bel-1961-book}
{\sc R.~Bellman}, {\em A brief introduction to theta functions}, Dover Publications, 1961.

\bibitem{Ber-1974-PRD}
{\sc C.~W. Bernard}, {\em Feynman rules for gauge theories at finite temperature}, Physical Review D, 9 (1974), pp.~3312--3320.

\bibitem{BjoDre-1965-book}
{\sc J.~Bjorken and S.~Drell}, {\em Relativistic quantum fields}, McGraw-Hill, New York, 1965.

\bibitem{BurFieHorSpe-1997-PRL}
{\sc D.~L. Burke, R.~C. Field, G.~Horton-Smith, J.~E. Spencer, D.~Walz, S.~C. Berridge, W.~M. Bugg, K.~Shmakov, A.~W. Weidemann, C.~Bula, K.~T. McDonald, E.~J. Prebys, C.~Bamber, S.~J. Boege, T.~Koffas, T.~Kotseroglou, A.~C. Melissinos, D.~D. Meyerhofer, D.~A. Reis, and W.~Ragg}, {\em Positron production in multiphoton light-by-light scattering}, Physical Review Letters, 79 (1997), pp.~1626--1629.

\bibitem{Con-1972-NPB}
{\sc D.~Constantinescu}, {\em Vacuum polarization in magnetic field}, Nuclear Physics B, 36 (1972), pp.~121--129.

\bibitem{Das-2011-JPCS}
{\sc A.~Das}, {\em Finite temperature effective actions}, Journal of Physics: Conference Series, 287 (2011).

\bibitem{DasFre-2009-PRD}
{\sc A.~Das and J.~Frenkel}, {\em Effective actions at finite temperature}, Physical Review D, 80 (2009), p.~125039.

\bibitem{DasFre-2010-PRD}
\leavevmode\vrule height 2pt depth -1.6pt width 23pt, {\em Thermal effective action for 1+1 dimensional massive {Q}{E}{D}}, Physical Review D, 82 (2010), p.~125002.

\bibitem{DasFre-2011-PLB}
\leavevmode\vrule height 2pt depth -1.6pt width 23pt, {\em The thermal chiral anomaly in the {S}chwinger model}, Physics Letters B, 704 (2011), pp.~85--88.

\bibitem{DenSve-2003-AA}
{\sc V.~I. Denisov and S.~I. Svertilov}, {\em Vacuum nonlinear electrodynamics curvature of photon trajectories in pulsars and magnetars}, Astronomy \& Astrophysics, 399 (2003), pp.~L39--L42.

\bibitem{Dir-1930-PRSL}
{\sc P.~A.~M. Dirac}, {\em A theory of electrons and protons}, Proceedings of the Royal Society of London. Series A, Containing Papers of a Mathematical and Physical Character, 126 (1930), pp.~360--365.

\bibitem{Dir-1934-MPCPS}
\leavevmode\vrule height 2pt depth -1.6pt width 23pt, {\em Discussion of the infinite distribution of electrons in the theory of the positron}, Mathematical Proceedings of the Cambridge Philosophical Society, 30 (1934), pp.~150--163.

\bibitem{Dir-1934-SR}
\leavevmode\vrule height 2pt depth -1.6pt width 23pt, {\em Théorie du positron}, Solvay report, Gauthier-Villars XXV, 353S,  (1934), pp.~203--212.

\bibitem{Dit-1979-PRD}
{\sc W.~Dittrich}, {\em Effective {L}agrangians at finite temperature}, Physical Review D, 19 (1979), pp.~2385--2390.

\bibitem{DitGie-2000-book}
{\sc W.~Dittrich and H.~Gies}, {\em Probing the quantum vacuum. Pertubative effective action approach in quantum electrodynamics and its application}, Springer Berlin, Heidelberg, 2000.

\bibitem{DolJac-1974-PRD}
{\sc L.~Dolan and R.~Jackiw}, {\em Symmetry behavior at finite temperature}, Physical Review D, 9 (1974), pp.~3320--3341.

\bibitem{Dun-2005-book}
{\sc G.~V. Dunne}, {\em Heisenberg-Euler effective lagrangians: basics and extensions}, World Scientific, 2005, pp.~445--522.

\bibitem{ElmPerSka-1993-PRL}
{\sc P.~Elmfors, D.~Persson, and B.-S. Skagerstam}, {\em {QED} effective action at finite temperature and density}, Physical Review Letters, 71 (1993), pp.~480--483.

\bibitem{ElmSka-1995-PLB}
{\sc P.~Elmfors and B.-S. Skagerstam}, {\em Electromagnetic fields in a thermal background}, Physics Letters B, 348 (1995), pp.~141--148.

\bibitem{EstLewSer-2008-BAMS}
{\sc M.~J. Esteban, M.~Lewin, and E.~S\'{e}r\'{e}}, {\em Variational methods in relativistic quantum mechanics}, Bulletin of the American Mathematical Society, 45 (2008), pp.~535--593.

\bibitem{EulKoc-1935-DN}
{\sc H.~Euler and B.~Kockel}, {\em Über die {S}treuung von {L}icht an {L}icht nach der {D}iracschen {T}heorie}, Die Naturwissenschaften, 23 (1935), pp.~246--247.

\bibitem{FanKamInaYam-2017-EPJD}
{\sc X.~Fan, S.~Kamioka, T.~Inada, T.~Yamazaki, T.~Namba, S.~Asai, J.~Omachi, K.~Yoshioka, M.~Kuwata-Gonokami, A.~Matsuo, K.~Kawaguchi, K.~Kindo, and H.~Nojiri}, {\em The {OVAL} experiment: a new experiment to measure vacuum magnetic birefringence using high repetition pulsed magnets}, The European Physical Journal D, 71 (2017).

\bibitem{FetWal-1971-book}
{\sc A.~L. Fetter and J.~D. Walecka}, {\em Quantum theory of many-particle systems}, International series in pure and applied physics, McGraw-Hill, New York, NY, 1971.

\bibitem{Fur-1937-PR}
{\sc W.~H. Furry}, {\em A symmetry theorem in the positron theory}, Physical Review, 51 (1937), pp.~125--129.

\bibitem{GonKouSer-2023-AHP}
{\sc D.~Gontier, A.~E.~K. Kouande, and E.~Séré}, {\em Phase transition in the {P}eierls model for polyacetylene}, Annales Henri Poincaré, 24 (2023), pp.~3945--3966.

\bibitem{GraRyz-1965-book}
{\sc I.~S. Gradshteyn and I.~M. Ryzhik}, {\em Table of integrals, series, and products}, Academic Press, New York, 1965.

\bibitem{GraHaiLewSer-2013-ARMA}
{\sc P.~Gravejat, C.~Hainzl, M.~Lewin, and E.~S\'{e}r\'{e}}, {\em Construction of the {P}auli-{V}illars-regulated {D}irac vacuum in electromagnetic fields}, Archive for Rational Mechanics and Analysis, 208 (2013), pp.~603--665.

\bibitem{GraHaiLewSer-2012-JEDP}
{\sc P.~Gravejat, C.~Hainzl, M.~Lewin, and E.~Séré}, {\em Two {H}artree-{F}ock models for the vacuum polarization}, Journées équations aux dérivées partielles,  (2012), pp.~1--31.

\bibitem{GraLewSer-2009-CMP}
{\sc P.~Gravejat, M.~Lewin, and E.~S\'{e}r\'{e}}, {\em Ground state and charge renormalization in a nonlinear model of relativistic atoms}, Communications in Mathematical Physics, 286 (2009), pp.~179--215.

\bibitem{GraLewSer-2011-CMP}
\leavevmode\vrule height 2pt depth -1.6pt width 23pt, {\em Renormalization and asymptotic expansion of {D}irac's polarized vacuum}, Communications in Mathematical Physics, 306 (2011), pp.~1--33.

\bibitem{GraLewSer-2018-JMPA}
\leavevmode\vrule height 2pt depth -1.6pt width 23pt, {\em Derivation of the magnetic {E}uler{\textendash}{H}eisenberg energy}, Journal de Math{\'{e}}matiques Pures et Appliqu{\'{e}}es, 117 (2018), pp.~59--93.

\bibitem{GreRei-2009-book}
{\sc W.~Greiner and J.~Reinhardt}, {\em Quantum electrodynamics}, Springer Berlin Heidelberg, 2009.

\bibitem{HaiLewSei-2008-RMP}
{\sc C.~Hainzl, M.~Lewin, and R.~Seiringer}, {\em A nonlinear model for relativistic electrons at positive temperature}, Reviews in Mathematical Physics, 20 (2008), pp.~1283--1307.

\bibitem{HaiLewSer-2005-JPA}
{\sc C.~Hainzl, M.~Lewin, and E.~S\'{e}r\'{e}}, {\em Self-consistent solution for the polarized vacuum in a no-photon {QED} model}, Journal of Physics A, 38 (2005), pp.~4483--4499.

\bibitem{HaiLewSol-2007-CPAM}
{\sc C.~Hainzl, M.~Lewin, and J.~P. Solovej}, {\em The mean-field approximation in quantum electrodynamics: The no-photon case}, Communications on Pure and Applied Mathematics, 60 (2007), pp.~546--596.

\bibitem{HaiSie-2003-CMP}
{\sc C.~Hainzl and H.~Siedentop}, {\em Non-perturbative mass and charge renormalization in relativistic no-photon quantum electrodynamics}, Communications in Mathematical Physics, 243 (2003), pp.~241--260.

\bibitem{HeiEul-1936-ZFP}
{\sc W.~Heisenberg and H.~Euler}, {\em Folgerungen aus der {D}iracschen {T}heorie des {P}ositrons}, Zeitschrift für Physik, 98 (1936), pp.~714--732.

\bibitem{KirLin-1972-PLB}
{\sc D.~Kirzhnits and A.~Linde}, {\em Macroscopic consequences of the {W}einberg model}, Physics Letters B, 42 (1972), pp.~471--474.

\bibitem{Lei-1975-RMP}
{\sc G.~Leibbrandt}, {\em Introduction to the technique of dimensional regularization}, Reviews of Modern Physics, 47 (1975), pp.~849--876.

\bibitem{MarBroSte-2003-PRL}
{\sc M.~Marklund, G.~Brodin, and L.~Stenflo}, {\em Electromagnetic wave collapse in a radiation background}, Physical Review Letters, 91 (2003), p.~163601.

\bibitem{MelSto-1976-INCA}
{\sc D.~B. Melrose and R.~J. Stoneham}, {\em Vacuum polarization and photon propagation in a magnetic field}, Il Nuovo Cimento A, 32 (1976), pp.~435--447.

\bibitem{MigTestGonTav-2016-MNRAS}
{\sc R.~P. Mignani, V.~Testa, D.~Gonzalez~Caniulef, R.~Taverna, R.~Turolla, S.~Zane, and K.~Wu}, {\em Evidence for vacuum birefringence from the first optical-polarimetry measurement of the isolated neutron star {RX} {J}1856.5-3754}, Monthly Notices of the Royal Astronomical Society, 465 (2016), pp.~492--500.

\bibitem{MouTajBul-2006-RMP}
{\sc G.~A. Mourou, T.~Tajima, and S.~V. Bulanov}, {\em Optics in the relativistic regime}, Reviews of Modern Physics, 78 (2006), pp.~309--371.

\bibitem{NikOuv-1983-book}
{\sc A.~Nikiforov and V.~Ouvarov}, {\em Fonctions spéciales de la physique mathématique}, Editions MIR Moscou, 1983.

\bibitem{PauVil-1949-RMP}
{\sc W.~Pauli and F.~Villars}, {\em On the invariant regularization in relativistic quantum theory}, Reviews of Modern Physics, 21 (1949), pp.~434--444.

\bibitem{ReeSim-1972-book}
{\sc M.~Reed and B.~Simon}, {\em Methods of modern mathematical physics. I. Functional analysis}, Academic Press, New York, 1972.

\bibitem{Sch-2013-book}
{\sc M.~D. Schwartz}, {\em Quantum field theory and the standard model}, Cambridge University Press, 2013.

\bibitem{Sch-1951-PR}
{\sc J.~Schwinger}, {\em On gauge invariance and vacuum polarization}, Physical Review, 82 (1951), pp.~664--679.

\bibitem{Ser-1935-PR}
{\sc R.~Serber}, {\em Linear modifications in the {M}axwell field equations}, Physical Review, 48 (1935), pp.~49--54.

\bibitem{Tha-1992-book}
{\sc B.~Thaller}, {\em The {D}irac equation}, Texts and Monographs in Physics, Springer-Verlag, Berlin, 1992.

\bibitem{TsaErb-1974-PRD}
{\sc W.-Y. Tsai and T.~Erber}, {\em Photon pair creation in intense magnetic fields}, Physical Review D, 10 (1974), pp.~492--499.

\bibitem{TsaErb-1975-PRD}
\leavevmode\vrule height 2pt depth -1.6pt width 23pt, {\em Propagation of photons in homogeneous magnetic fields: Index of refraction}, Physical Review D, 12 (1975), pp.~1132--1137.

\bibitem{Ueh-1935-PR}
{\sc E.~A. Uehling}, {\em Polarization effects in the positron theory}, Physical Review, 48 (1935), pp.~55--63.

\bibitem{Wei-1974-PRD}
{\sc S.~Weinberg}, {\em Gauge and global symmetries at high temperature}, Physical Review D, 9 (1974), pp.~3357--3378.

\bibitem{Wei-1936-MFM}
{\sc V.~Weisskopf}, {\em Über die {E}lektrodynamik des {V}akuums auf {G}rund der {Q}uantentheorie des {E}lektrons}, Mathematisk-fysiske Meddelelser, 16 (1936), pp.~1--39.

\end{thebibliography}

\end{document}